\documentclass[12pt,preprint]{aastex}
\usepackage{emulateapj5}
\shorttitle{Variable Stars in $\omega$ Centauri}
\shortauthors{Weldrake, Sackett \& Bridges}

\begin{document}
\title{A Deep Wide-Field Variable Star Catalog of $\omega$ Centauri}

\author{David T F Weldrake} 
\affil{Max Planck Instit\"ut F\"ur Astronomie, K\"onigstuhl 17, D-69117, Heidelberg, Germany}
\email{weldrake@mpia-hd.mpg.de}

\author{Penny D Sackett}
\affil{Research School of Astronomy and Astrophysics, Australian National University, Mount Stromlo Observatory, Cotter Road, Weston Creek, ACT 2611, Australia}
\email{Penny.Sackett@anu.edu.au}

\author{Terry J Bridges}
\affil{Physics Department, Queen's University, Kingston, Ontario, Canada K7L 3N6}
\email{tjb@astro.queensu.ca}

\begin{abstract}
We present a variable star catalog of an extensive ground-based wide-field variability survey in the globular cluster $\omega$ Centauri. Using the ANU 40-inch (1m) telescope at Siding Spring Observatory, the cluster was observed with a 52$'$$\times$52$'$ (0.75 deg$^2$) field for 25 nights. A total of 187 variable stars were identified in the field, 81 of which are new discoveries. This work comprises the widest field variability survey yet undertaken for this cluster. Here we present the V+R lightcurves and preliminary analysis of the detected variable stars, comprising 58 eclipsing binaries, 69 RR Lyrae stars, 36 long period variables (P$\ge$2d) and 24 miscellaneous pulsators including 15 SX Phoenicis stars and two Type II Cepheids. 

Analysis of the eclipsing binary radial distribution has revealed an apparent lack of binaries in the 8$'$-15$'$ range, perhaps indicating two separate binary populations. Four detached binaries have short periods ($<$2.5d) and are likely composed of low-mass M-dwarf components, useful for testing stellar evolution models. One further detached system has a period of 0.8 days and due to the blueness of the system could be composed of white dwarf stars. Analysis of the RR Lyrae sample has produced a reddening corrected distance modulus (also accounting for metallicity spread) for the cluster of 13.68$\pm$0.27, a result consistent with previously published values. This paper also presents a total stellar database comprising V and I photometry (with astrometry better than 0.25$''$) for 203,892 stars with 12.0$\leqslant$V$\leqslant$21.0 and 25-night V+R lightcurves for 109,726 stars (14.0$\leqslant$V$\leqslant$22.0) for both the cluster and the field.
\end{abstract}

\keywords{globular clusters: individual $\omega$ Centauri (NGC 5139) --- binaries: eclipsing --- binaries: general --- stars: variables: Delta Scuti --- other}

\section{Introduction}
Omega Centauri ($\omega$ Cen) has been the subject of intense interest over the years, as it possesses several distinctive features that differentiate it significantly from other members of the Galactic globular cluster system. Firstly, it is the most massive of the globular clusters, with a total absolute visual magnitude of $-$10.29 \citep{Harris96}, comparable to low-mass dwarf galaxies. The cluster posesses a high internal rotation velocity of $\sim$8 km/s$^{-1}$, \citep{Merr1997} with a central one-dimensional velocity dispersion of $\sim$15$-$20 km/s$^{-1}$ \citep{MM86,F2001,VDV2005}. This high rotation and mass (2.5$\pm$0.3$\times$10$^{6}$M${_{\odot}}$, \citet{VDV2005}) gives the cluster a moderate ellipticity and a long relaxation time \citep{DJ1993}. An investigation into the global dynamics of the cluster can be found in \citet{VDV2005}.

Perhaps most importantly, $\omega$ Cen is well known to display a complex stellar population, with a distinct metallicity spread among its stars \citep{DW1967,NB75,L1999,P2000,S2005}. This indicates that the cluster has undergone a star formation and chemical enrichment process that has been occuring over an extended period of time. Using helium abundances, \citet{N2004} has shown that the cluster has three distinct stellar populations, with corresponding metallicities ranging from $-$1.7 to $-$0.6 dex, with the majority of the stars (80$\%$) belonging to the metal-poor branch. These values correspond to an age spread of around 2-3 Gyr. Furthermore, only the metal-poor populations ([Fe/H]$\le$$-$1.2) seem to show evidence of rotation \citep{No1997,Xie2002}, although \citet{VDV2005} find no significant difference between the two populations. All observations suggest that the populations have different dynamical origins.

The origin of the cluster has been the subject of much debate, with the suggestion by several groups that it is the remnant of a dwarf galaxy disrupted by the Milky Way \citep{BF2003,IM2004}. Indeed, \citet{BN2005} postulate that the second generation population of $\omega$ Cen could have been formed from gas ejected from primordial stars which surrounded the cluster when it was once the nucleus of a dwarf galaxy. 

The lack of observed mass segregation in the cluster \citep{A1997,DSR2005,Feretal2006} may be further evidence that $\omega$ Cen was orginally a more massive object. Based on the current mass of $\omega$ Cen and a distance of 5.5Kpc, \citet{Feretal2006} calculate a central relaxation time of $\sim$6.6 Gyr, approximately a factor of two smaller than the cluster age, and thus expect there to be some observable mass segregation in the cluster center. However, \citet{Feretal2006} do not find any segregation in the cluster blue stragglers. They suggest two possible resolutions: that $\omega$ Cen was originally a much larger dwarf galaxy with a correspondingly larger relaxation time, and/or that the large rotation in $\omega$ Cen has increased the relaxation time, since angular momentum will keep stars out of the core. Further evidence of an external origin is that the cluster follows a highly bound retrograde orbit \citep{D1999}. 

One way in which the cluster dynamical evolution can be studied is via variable stars, particularly eclipsing binaries. Detached binaries located at the cluster main sequence turnoff allow direct determination of the properties of turnoff stars in the cluster, important for the verification (or otherwise) of theoretical cluster isochrones. The detection of any new detached M-dwarf binaries also allows comparison with models of stellar evolution \citep{Rib2000}. Binaries also provide calculations of stellar masses, radii, ages, luminosities \citep{Gim2001} and information on the evolution of both contact and detached systems. Hence, detection of new variable stars (particularly in the relatively understudied outer halo of the cluster) is useful for multiple scientific goals in furthering the understanding of the $\omega$ Cen stellar content.

Previous ground-based searches for cluster variable stars centered on the core have uncovered a rich population of many different types of variables, including SX Phe stars, eclipsing binaries (detached, semi-detached and contact systems), many RR Lyrae, spotted variables (variability associated with the rotational modulation of large starspots) and long period variables. The OGLE project has been the most prolific of these searches, with a total of 394 variables in their online catalogs \citep{Kal96,Kal97,Kal97b,K2004}. Other deep variability searches have been more limited to the cluster core, for example \citet{Hag2002}. 

The work presented here constitutes the results of a 25-night search for variability in a wide field centered on $\omega$ Cen which extends further from the cluster core and to deeper photometry than any previous ground-based variability survey.  The search extends to $\sim$50$\%$ of the cluster tidal radius (in a single exposure), corresponding to $\sim$6.8 times the cluster half mass radius \citep{Harris96}. Hence the observed field contains a large majority of the bound cluster stars and has excellent prospects for the discovery of new variable systems, both in the cluster and the field.

The main motivation for the observations is to search for transiting `Hot Jupiter' planets in the cluster, and is the second part of a search for such planets in both 47 Tucanae \citep{W2005} and $\omega$ Cen. These two clusters are unique in that they display sufficient star brightness and total star numbers for meaningful statistics to be gained from a ground-based campaign with a telescope of moderate aperture. The results of the Hot Jupiter search in $\omega$ Cen will be published in a separate paper \citep{W2006}. 

Section 2 of this work describes the observations and data reduction techniques employed to produce the total image dataset. Section 3 presents a description of the method used for time-series production and a discussion of the resultant photometric precision. Section 4 describes the cluster Color-Magnitude dataset and astrometry, along with a description of the theoretical isochrones produced for the cluster. Section 5 details the methods used to detect the variable stars and section 6 presents the variable star catalog itself, with descriptions of the color-magnitude distribution of the variables, their spatial distributions and a discussion of the resulting catalog detection limits. Section 7 describes the analysis of individual variables (eclipsing binaries, RR Lyrae and miscellaneous pulsators), and Section 8 presents the paper summary and conclusions.

\section{Observations and Data Reduction}
The image dataset was obtained using the Australian National University (ANU) 40-inch (1m) telescope located at Siding Spring Observatory, fitted with the Wide Field Imager (WFI). This telescope and detector combination permits a 52$'$$\times$52$'$ (0.75 deg${^2}$) field of view, capable of observing a large fraction of the cluster with a single exposure, thus maximising the number of sampled stars for lightcurve production. WFI consists of a 4$\times$2 array of 2048$\times$4096 pixel back-illuminated CCDs, arranged to produce a total array of 8K $\times$ 8K pixels. The detector scale is 0$^{``}$.38 pixel$^{-1}$ at the 1m telescope Cassegrain focus, allowing for suitable sampling of the point spread function (PSF) with the seeing limitations of the site. Our exposure times were fixed at 300 seconds resulting in excellent temporal resolution in the dataset. 

The main aim of the project is the detection of `Hot Jupiter' planet transits against $\omega$ Cen main sequence stars, which requires a signal-to-photon-noise ratio (S/N) of 200 or more for sufficient photometric precision ($\sim$1.5$\%$) at V$=$18.0 (typical V magnitude of the target stars in these crowded fields). This requirement therefore defined the observing strategy. In order to achieve this with short exposure times, a broadband V+R filter was used, covering the combined wavelength range of the Cousins V and R filters. This same telescope and detector combination was used for a deep search for planet transits in the halo of 47 Tucanae, yielding a high significance null result \citep{W2005}. A total of 69 newly discovered variable stars were also found in that search \citep{W2004}. From this previous experience a star of V$=$18.5 in 2$''$ seeing (typical of the site) yields a photon noise S/N of 220 with a 7-day moon and 165 at times of bright moon in a five minute exposure.

$\omega$ Cen was observed for a total of 25 contiguous nights, from 2003 May 2 to 2003 May 27 with the field centered at R.A $=$ 13$^{h}$26$^{m}$45.89$^{s}$, decl. $=$ $-$47$^{\circ}$28$'$36.7$''$. The position of the centers of all CCDs can be found in Table\space1. A total of 875 images of the cluster were obtained during this time, with an average temporal resolution of 6 minutes and covering an average 9 hours for each good night. Each image was independently checked at the telescope for cosmetic quality, and any with bad seeing ($>$3.5$''$ images), satellite trails or other adverse effects were discarded from the dataset. Of this total database, 90$\%$ were deemed useful for the analysis, having suitable seeing and small telescope offsets to minimise star loss. A total of 787 images (with average seeing of 2.1$''$) were subsequently used to produce 109,726 time-series via Differential Imaging Analysis (DIA, \citet{Woz2000}) across the whole field for a detailed variability analysis. 

Initial image reduction was performed according to standard practise within the MSCRED package of IRAF.\footnote{IRAF is distributed by National Optical Observatories, which is operated by the Association of Universities for Research in Astronomy, Inc., under cooperative agreement with the National Science Foundation.} This incorporated region trimming, overscan correction, bias correction, flat-field and dark current subtraction. The images were then checked for flatness and overall quality and became available for the main photometric analysis and time-series production.

\section{Photometry and Photometric Accuracy}
In order to obtain high precision photometry on relatively faint targets in the crowded field of a globular cluster, differential photometry was performed on the dataset. This method was originally described as an optimal Point-Spread-Function (PSF) matching algorithm by \citet{AL98}, and was subsequently modified by \citet{Woz2000} for use in detecting microlensing events. A detailed description of the method and software pipeline can be found in Wozniak's paper and only the main steps shall be summarized here. 

The process of matching the stellar PSF throughout a large database of images dramatically reduces the systematic effects of varying atmospheric conditions on resultant photometric precision, allowing ground-based observations the best chance of detecting small brightness variations in faint targets. DIA is also one of the optimal photometric methods for dealing with crowded fields, as a larger number of pixels contain information on any PSF differences as the number of stars increases, hence improving the PSF matching process. Flux measurements of the stars are made via profile photometry on a master template frame, produced via the median-combining of a large number of the best quality images with small offsets. This template image is used as the zero-point in the output time-series. 

The positions of the stars are found on a reference image, usually the image with the best seeing conditions, and all subsequent images in the dataset, including the template, are shifted to match. The best PSF-matching kernel is then found, and each registered image is subtracted from the template, with the residual subtraction generally being dominated by photon noise. Any object that has changed in brightness between the image and the template is given away as a bright or dark spot. 

Differential photometry produces time-series measured in differential counts, a linear flux unit from which a constant reference flux (taken from the template) has been subtracted. In order to convert to a standard magnitude system, the total number of counts for each star was measured using the PSF photometry package of DAOPHOT within IRAF, with the same images and parameters as used in the photometry code. The time-series were then converted using these flux values into magnitude units via the relation:

\begin{center}
$\Delta m_i = -2.5 \log [(N_i+ N_{\rm ref,\it{i}}) / N_{\rm ref,\it{i}}]$
\end{center}

\noindent where $N_{\rm ref,\it{i}}$ is the total flux of star $\it{i}$ on the template image and $N_i$ is the original difference flux in the time series as produced with the photometric code. 

The pixel coordinates of all visible stars were determined separately from the reference frame via DAOFIND within IRAF, and the profile photometry was then extracted from the subtracted frames at those determined positions. In this way, a total of 109,726 stars were identified and their time-series produced, across the whole WFI field, which then became the subject of analysis. The time-series are hence presented in this work in V+R differential magnitude units, and can be converted to the standard V system via the calculation of color terms. This requires another set of observations which is not directly related to the paper at hand. 

\subsection{Photometry of the cluster core}
Each of the four outer CCDs of WFI were analysed in half CCD chunks, each half producing an average of 9,500 time-series. For the crowded core of the cluster, the number of stars becomes very large and, due to computational limitations, a different strategy was employed. For the core regions, the images were analysed individually with DIA in 360 individual subframes, 90 per CCD. The locations of these subframes were chosen so that no stars were lost at the edges of the subframes, covering the entire core region of the cluster, except those regions that were affected by telescope offsets during observing (a 160 pixel border surrounding each CCD). 

\begin{figure}
\plotone{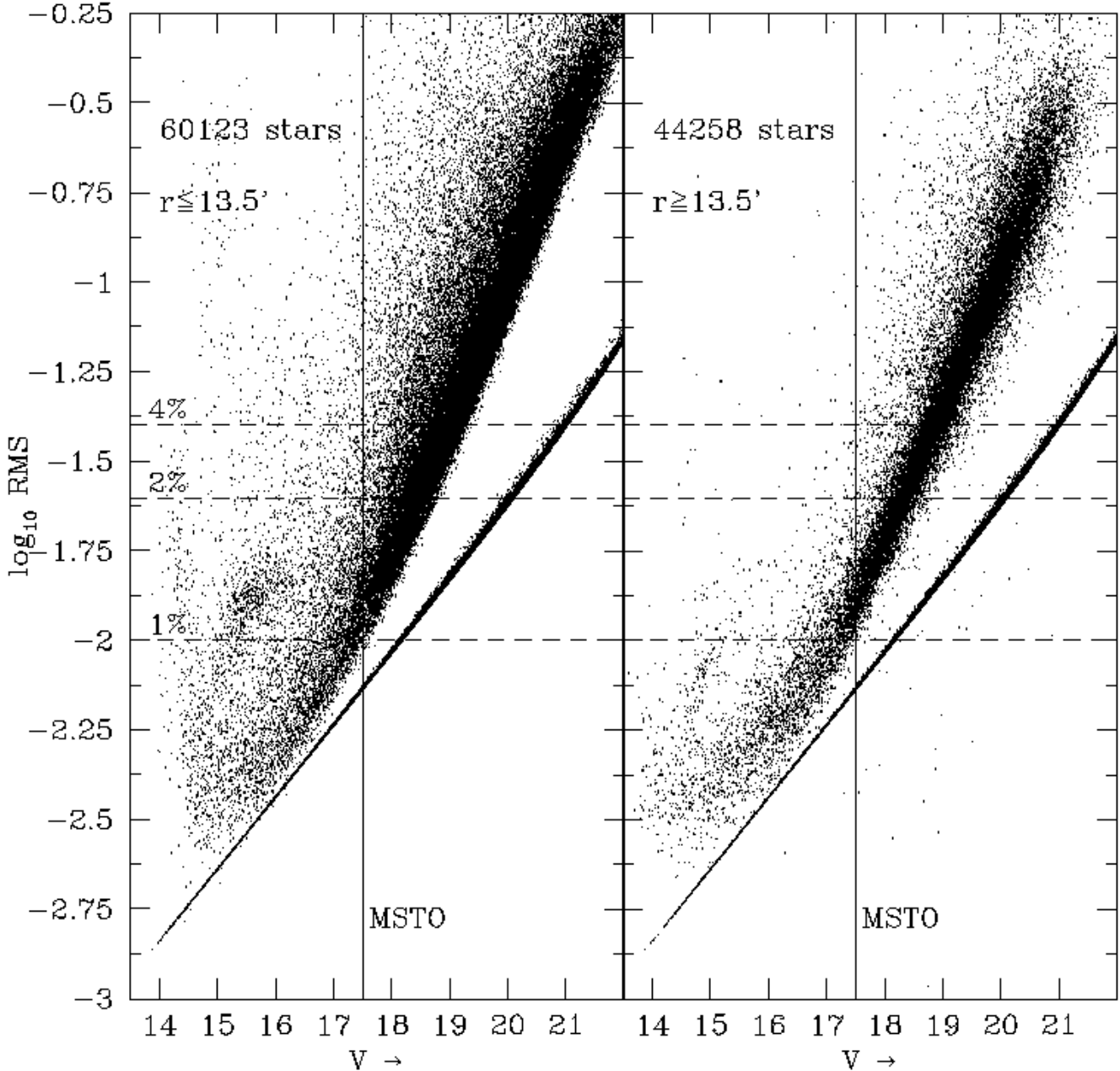}
\figcaption[figure1.eps]{The measured DIA photometric scatter for a total of 104,381 stars which were cross-identified with the CMD dataset. The mean total star+background photon noise contribution is also indicated for each star (thin locus of points). Those stars located within and outside 13.5$'$ of the cluster core are plotted separately; the difference in photometric precision between these two crowding regimes is marginal. The position of the cluster main sequence turnoff (MSTO) is marked to indicate photometric rms for cluster main sequence stars. The uncertainty is 1$\%$ at the MSTO and 4$\%$ (0.04 mag) at V$\sim$19.0.\label{rmsplot}}
\end{figure}

Fig.\space\ref{rmsplot} presents the resultant DIA-derived photometric precision, measured as root-mean-square scatter (rms), for a total of 104,381 stars which were cross-identified with the cluster CMD dataset. The left panel displays the rms of 60,123 stars within 13.5$'$ of the cluster core and the right panel shows the rms of 44,258 stars outside this radius. The difference between these two crowding regimes is marginal, indicating the ability of DIA to handle fields of differing stellar density. The mean total star and background noise contribution is also plotted, illustrating that the photometry is photon-noise-limited to V$\sim$17.0. 

The position of the cluster main sequence turnoff (MSTO) is marked to indicate where the cluster stars become members of the main sequence. To the left of this line, therefore, lie likely Red Giant Branch and foreground Galactic disk stars. It can be seen that the photometric accuracy is equal to 2$\%$ (0.02 magnitudes) at V$\sim$18.5 for both crowding regions of the data, increasing to 4$\%$ (0.04 magnitudes) at V$\sim$19.0. By considering the stellar radius for cluster main sequence stars as a function of V, the photometry allows detection of transiting giant planets down to V$\sim$19.5.

\section{Color Magnitude Diagram and Astrometry}
\begin{figure}
\plotone{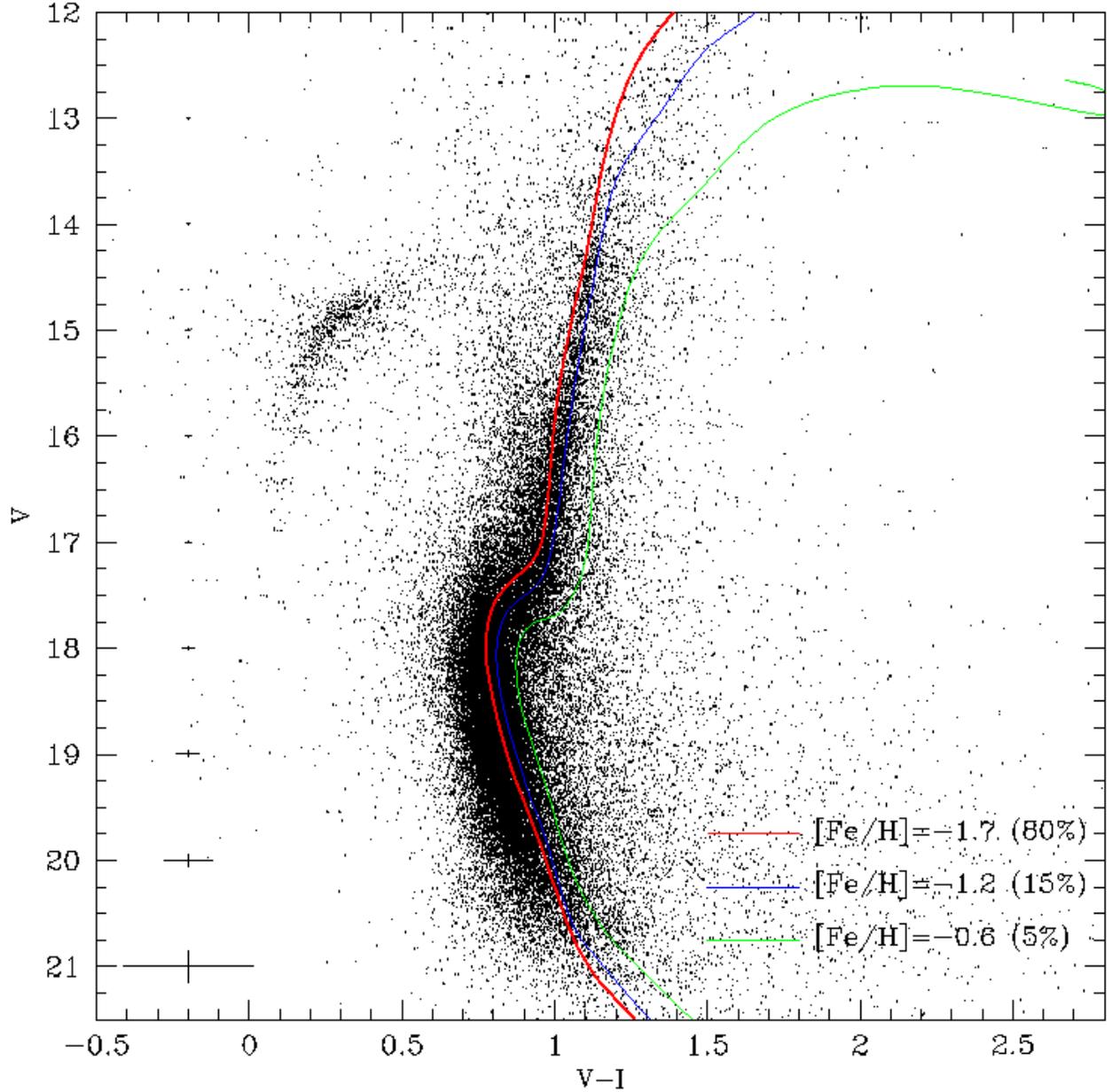}
\figcaption[figure2.eps]{The Color Magnitude Diagram (CMD) dataset used to derive color information for variable objects. The DAOPHOT output errors are marked as errorbars as a function of V. Also overplotted are the three theoretical \citet{Y2003} isochrones to define the stellar populations of the cluster. The metallicity and relative fraction of each isochrone as taken from \citet{N2004} is marked accordingly. The CMD calibration is accurate to better than 0.03 mag as described in the text.\label{cmdplot}}
\end{figure}

Using the same telescope/detector combination and pointing, a V, V-I color magnitude diagram (CMD) totalling 203,892 stars was produced for the observed field. This enabled detected variable stars and transiting systems to be placed on the standard V and I magnitude system, aiding in the determination of their likely nature. Three images in V and three in I were combined to produce the dataset, with all images being taken within 30 minutes to minimise the effect of variability on the resulting magnitudes and colors. Fig.\space\ref{cmdplot} presents the diagram produced for all CCDs. The output DAOPHOT photometric errors in both V and V-I are marked as errorbars as a function of V magnitude. The magnitude range of this diagram (12.0$\le$V$\le$21.0) covers a large range of stellar mass both in the cluster and the contaminating Galactic disk. 

The CMD calibration was performed via matching of stellar astrometry from our catalog to that of \citet{C2004} (also taken with the ANU 1m and WFI combination in V and I), as standard field data were unavailable. The difference in V and I between our uncalibrated data and the \citet{C2004} calibrated data was measured for each of the matched stars (totalling more than 20,000) in each CCD independently; the resultant calibration accuracy was $\le$0.03 magnitudes. 

Also overplotted on Fig.\space\ref{cmdplot} are three theoretical \citet{Y2003} isochrones used to simulate the stellar populations of the cluster. These isochrones were used to determine stellar mass and radius values for cluster main sequence stars for use in the Hot Jupiter transit search. They also allow investigation into whether any particular type of variable is preferentially located within a particular population. The metallicity and relative fraction of each cluster population, as taken from \citet{N2004}, is also plotted.

Astrometry was obtained for a total of 212,959 stars identified in the V band image of the cluster, and 243,466 stars in the I-band. A search of the USNO CCD Astrograph Catalog (UCAC1) was carried out for astrometric standard stars within the field. Several hundred such stars were successfully identified, producing an accurate determination of the astrometric solution for the stars in each CCD, with measured uncertainties of 0.25$''$. 

\begin{figure}
\plotone{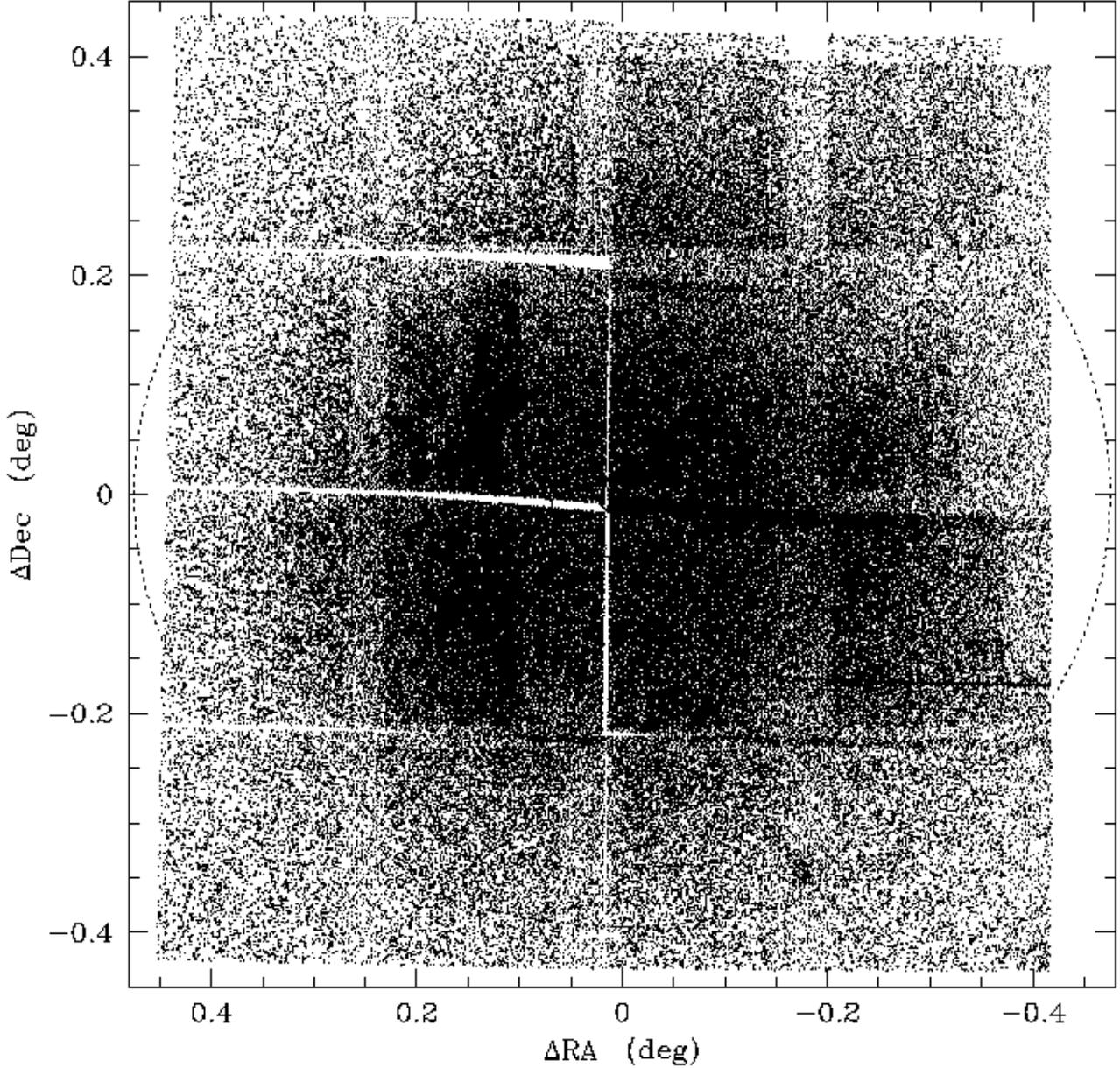}
\figcaption[figure3.eps]{The total derived V band astrometry (light shading) plotted as $\Delta$RA and $\Delta$Dec (in degrees) from the location of the cluster core for 212,957 stars. The extent of the eight WFI CCDs are clear, as well as the differing stellar densities encountered in the dataset. Also overplotted (darker shading) are the locations of stars for which time-series production was possible. The gaps in the spatial coverage are caused by the spaces separating CCDs and regions where stellar density (and saturation) did not allow sufficient accuracy in the output DIA templates and subtracted frames or by telescope offsets. We have time-series information for 54$\%$ of the stars for which we have astrometry. The three ellipses (plotted with the $\omega$ Cen ellipticity and cluster parameters taken from \citet{Harris96}) define the positions of the cluster core radius (inner ellipse), the cluster half-mass radius (central ellipse) and the position of 50$\%$ of the cluster tidal radius (outer ellipse) - the extent of our single WFI field of view.\label{clusterastrom}}
\end{figure}

In order to display the extent of our field of view with respect to the cluster, Fig.\space\ref{clusterastrom} presents both the total derived V band astrometric dataset (light shading) and the total time-series dataset (dark shading), plotted as $\Delta$RA and $\Delta$Dec in degrees from the location of the cluster core. The eight CCDs of WFI are clear, as is the differing stellar densities encountered in the dataset. The time-series database (overplotted with darker shading) does not have the completeness of the total astrometry. The gaps are in regions where poor photometry resulted due to the presence of bright saturated stars or repeated measurements could not be obtained in a border surrounding each chip due to telescope offsets during the run. A total of 109,726 lightcurves were produced in the sampled regions.

Also overplotted on Fig.\space\ref{clusterastrom} as ellipses are the locations of the cluster core radius (innermost ellipse), the cluster half-mass radius (central ellipse) and the position of 50$\%$ of the cluster tidal radius (outer ellipse, the extent of the search). The ellipses have been plotted with the $\omega$ Cen ellipticity and cluster parameters of \citet{Harris96}.

\begin{figure}
\plotone{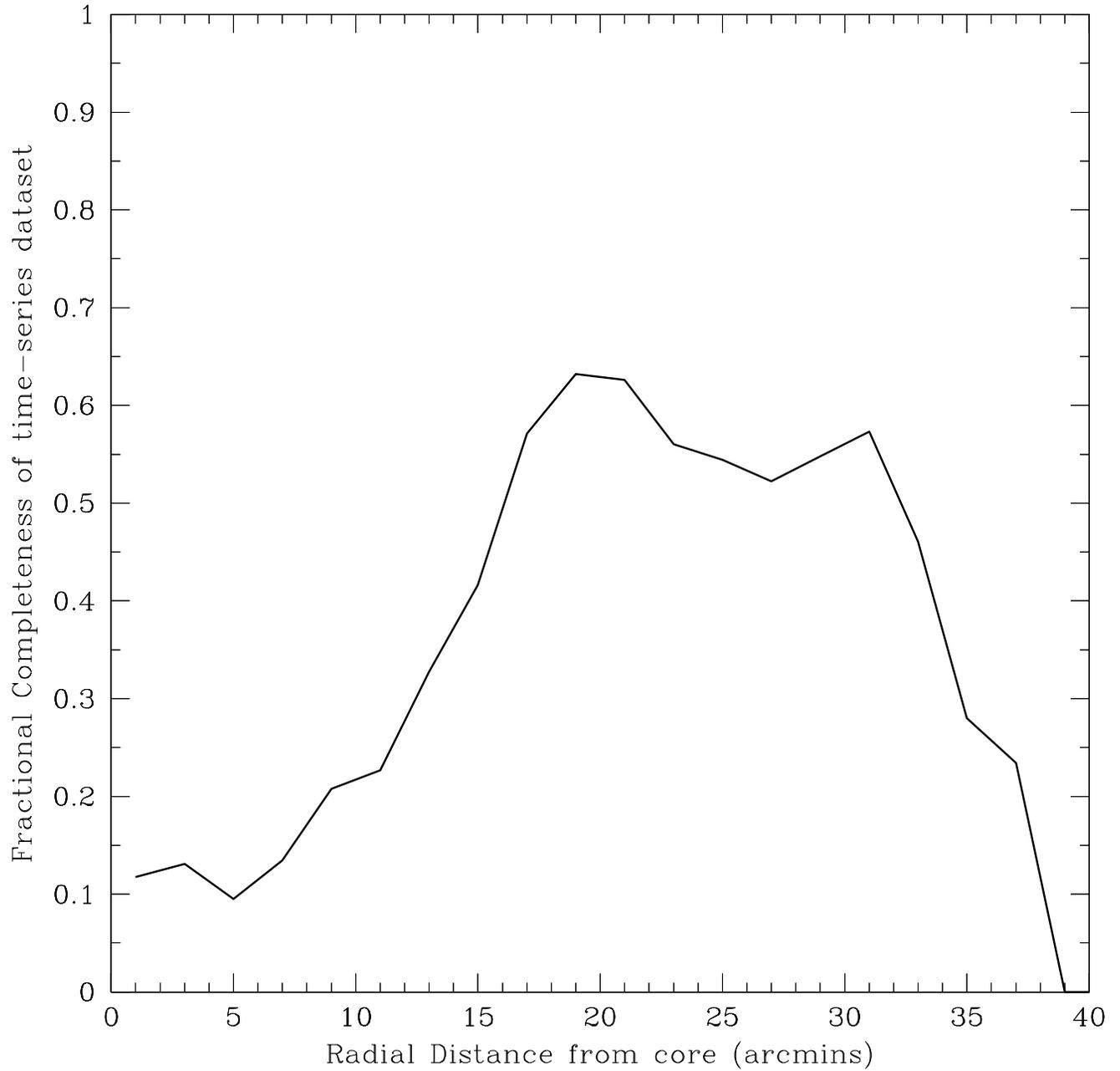}
\figcaption[figure4.eps]{The fraction of stars with time-series information compared to the total astrometric stellar database, as a function of radial distance from the cluster core. The truncation in fraction is caused by the gaps in spacial coverage of the cluster, as seen in Fig.\space\ref{clusterastrom}. The time-series dataset has an optimal zone from $\sim$18$'$ to 32$'$ from the cluster core.\label{completeness}}
\end{figure}

Fig.\space\ref{completeness} shows the fraction of stars for which time-series information is available compared to the total astrometric database, as a function of radial distance from the cluster core. Our time-series database has an optimal region of $\sim$18$'$ to $\sim$32$'$ from the core. The decrease in completeness in the core and in the outer regions of the field are due to the gaps in the spatial coverage as seen in Fig.\space\ref{clusterastrom}. 

\section{Variable Star Detection Methods}
In order to automatically detect the variable stars in the total time-series database, two search methods were used. For any search, two main factors must be taken into account, namely the distribution of the observations in time and the shape of the variability for which the search is targeted. First, we applied the Lomb-Scargle Periodogram (LSP) \citep{Brett01} in which a Fourier power spectrum is produced with the same statistical properties as standard power spectra, but successfully overcomes the problems caused by diurnal gaps. The method produces a spike (with a frequency of 2$\pi$/$\it{P}$) in the output power spectrum if a significant periodicity ($\it{P}$) is detected. The significance (in multiples of the standard deviation of the spectrum, $\sigma$) is determined for each datapoint independently. 

If any datapoint is over a set detection threshold, a variable star candidate is flagged. By experimentation, it was found that for our dataset setting this detection threshold at $\ge$12$\times$$\sigma$ produces a variable star recoverability of 1 real variable per $\sim$1,000 stars searched, with a corresponding false detection probability (per star) of 0.002$\%$. In this manner, all stars in the time-series database were searched and the first sample of variable stars identified. 

However, this method is only of benefit when searching for sinusoidal variability, and will miss detached systems and other non-sinusoidal stars. To overcome this, a second search was implemented on all stars using an application of the Analysis of Variance (AoV) statistic (O.Tamuz, private communication). A full description of this detection method can be found in \citet{S1989}. Via this method, the data are phase-wrapped with a trial period and then grouped into phase bins. A one-way statistical analysis of variance is then performed on the result. This statistical procedure is repeated for a fixed range of test periods for each star, producing a series of significances and their corresponding periodicities. The final output for each star is the peak periodicity and its corresponding significance. 

Period estimates for the detected variables were derived by measuring the position of the highest significance peak in either the output power spectrum or the peak AoV statistic periodicity (in day units to 4 decimal places) and phase-wrapping the star at that period. The period was then tweaked manually to produce the smallest amount of scatter on the plotted lightcurve. The difference between the raw measured period and the final plotted period was smaller than 0.001 day in all cases, an indication of the accuracy of the detection methods. 

\section{Variable Star Catalog Overview}
As a result of analysing the whole dataset of 109,726 lightcurves, a total of 530 candidates were produced with a significant ($\ge$12$\sigma$) periodicity as determined via LSP along with 2324 candidates with high significance ($\ge$8$\sigma$) as determined with AoV. The AoV candidates include all of those detected via LSP. All candidates were then examined by eye both in their `raw' un-phase-wrapped format and phase-wrapped to their peak detected periodicity. If the phase-wrapped lightcurve displayed discernable regular variability at the periodicities detected, they were flagged as variable stars. The vast majority of the candidates (particularly those identified by AoV) were found to be attributed to common systematic effects inherent to the data, associated with stars close to the magnitude limits of the dataset.

In all, a total of 187 secure variable stars were identified, across the whole WFI field and are presented in Table.\space2. The final catalog consists of 58 eclipsing binaries (EcB), 69 RR Lyrae stars, 36 Long Period Variables (LPVs, P$\ge$2d) and 24 miscellaneous variables including 15 SX Phoenicis ($\delta$ Scuti) stars. From their locations on the cluster CMD, most of these systems are expected to be cluster members. Follow-up radial velocity observations are needed to confirm memberships. Of this sample, 81 of these variables are new discoveries as indicated by cross-matching astrometry with the \citet{K2004} catalog. The time-series data are available on the electronic edition of AJ.

\subsection{Comparison with Previous Studies}
All of the detected variables were compared to the online catalog of \citet{K2004} in order to identify new discoveries. A comparison was made of the published astrometry, period, type and V magnitudes of the known variables to the corresponding values derived in this work. We were able to match 106 of our variables with those of \citet{K2004}.

\begin{figure}
\plotone{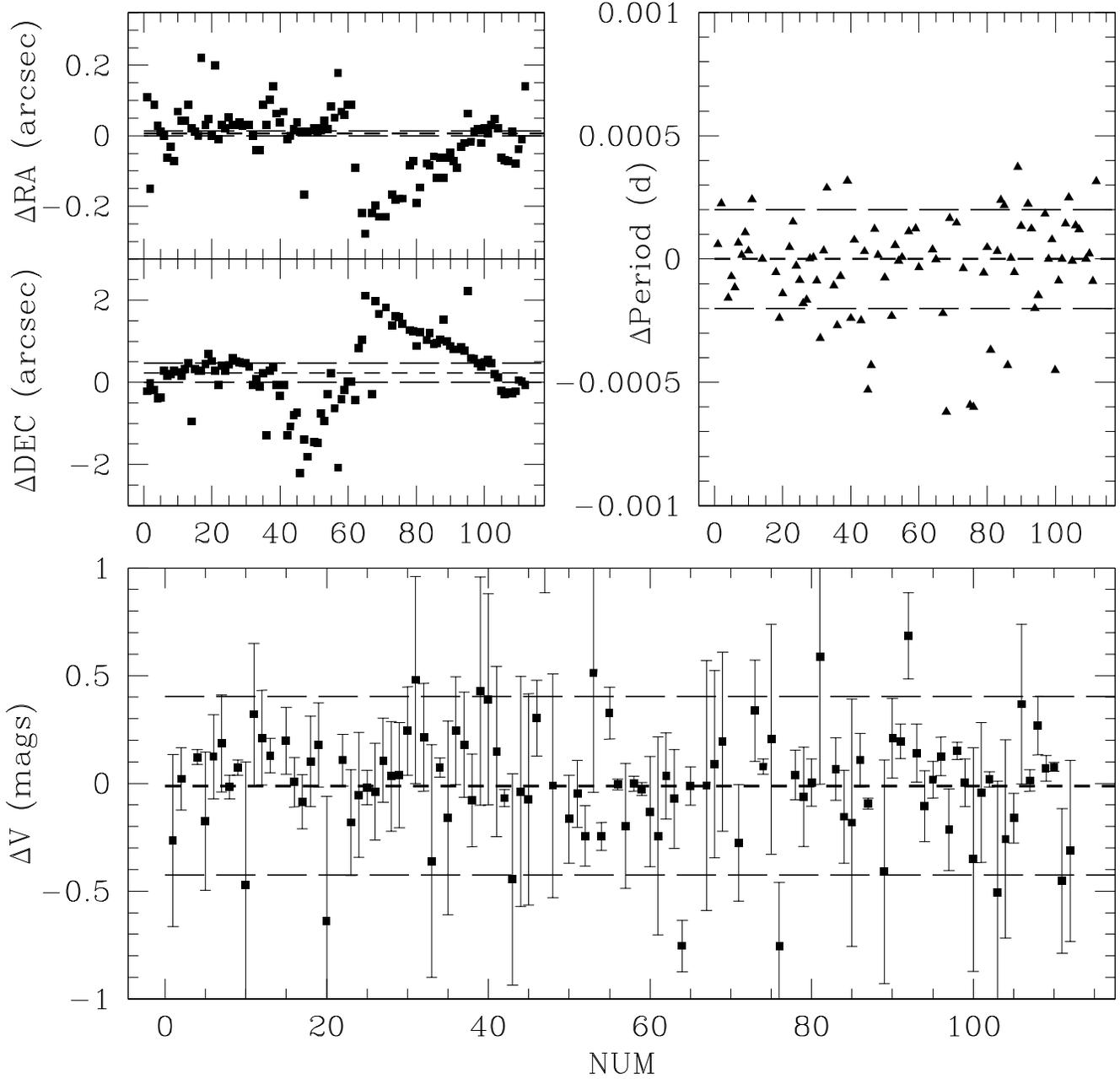}
\figcaption[figure5.eps]{A comparison of previously known variables from the \citet{K2004} online database, with an arbitrary identification number on the X-axis. Plotted are the differences in astrometry (top left panel), the differences in derived period (top right panel) and the differences in measured V magnitude (bottom panel). The zero-point for all panels has been overplotted for comparison. The difference in the astrometry for those crossidentified variables is less than RA$=$0.3$''$ and DEC$=$3$''$ for all cases and the derived periods are seemingly accurate between the two datasets to P$\sim$0.0005d. The V magnitude values presented in this work are measured directly from the CMD dataset, and have a subsequent scatter of $\sim$0.4 magnitudes when compared to the average V magnitudes of \citet{K2004}. The errorbars denote the range in V magnitude the variable undergoes. Our V magnitudes are within these errorbars for the majority of cases.\label{comparevars}}
\end{figure}

Fig.\space\ref{comparevars} shows the results of the comparisons made for these 106 recovered variables. The zeropoint was determined along with the standard deviation and has been overplotted for all parts of the figure for comparison. The comparisons are all consistent with zero. All matches within three arcseconds in both RA and DEC were classified as recovered variables, with the difference in derived astrometry both for RA and DEC seen in the top left panel of the figure. The final matching threshold was chosen by varying threshold, and comparing the number and period determinations of matches between the two datasets. Three arcseconds was found to produce the largest number of matches with the periods being consistent, a larger threshold introduces mis-matches. The average of the astrometry differences was found to be 0.005$\pm$0.005$''$ for the RA astrometry and 0.232$\pm$0.233$''$ for DEC. There is a slight systematic offset of 0.23$''$ in declination between our catalog and that of \citet{K2004}. The sample of stars for which a large offset in astrometry is seen (num 40-60) is due to those variables being located in the most crowded regions of the cluster.

The top right panel shows the difference in derived period for these 106 stars. All recovered variables except six have periods presented here within 0.0005d of those published by \citet{K2004}, with an average very close to zero and standard deviation of 0.0002d. The remainder are all long period variables with a larger error in their period determination, due to incomplete phase coverage.

The bottom panel of Fig.\space\ref{comparevars} shows the difference in V magnitude as determined from our CMD dataset, and the average V magnitude as determined by \citet{K2004}. The plot shows some scatter around the zeropoint with an amplitude $\sim$0.5 magnitudes, and has an average variation of -0.01$\pm$0.42 mags. The range in V magnitude undertaken by the variable during the course of it's variation is plotted as an errorbar for each point. Our measured V magnitude zero-point is within these errobars for the vast majority of cases, indicating that the difference in V between our dataset and that of \citet{K2004} (and the subsequent 0.42 magnitude error) is caused by the various phases that each variable was undertaking when our CMD dataset was obtained.

\begin{figure}
\plotone{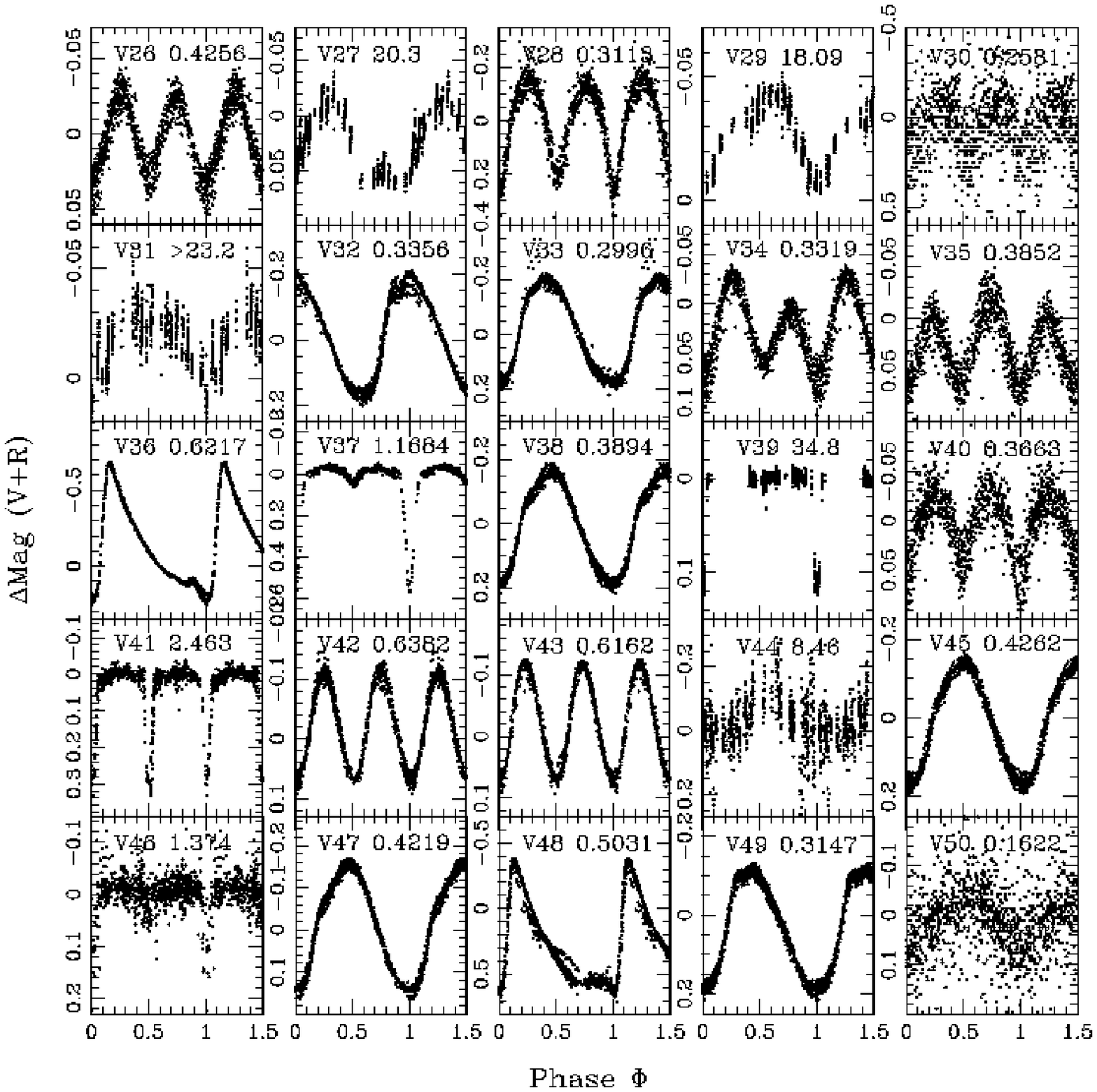}
\figcaption[figure10.eps]{Phase-wrapped differential V+R magnitude variable star lightcurves (continued).\label{varplot2}}
\end{figure}

\begin{figure}
\plotone{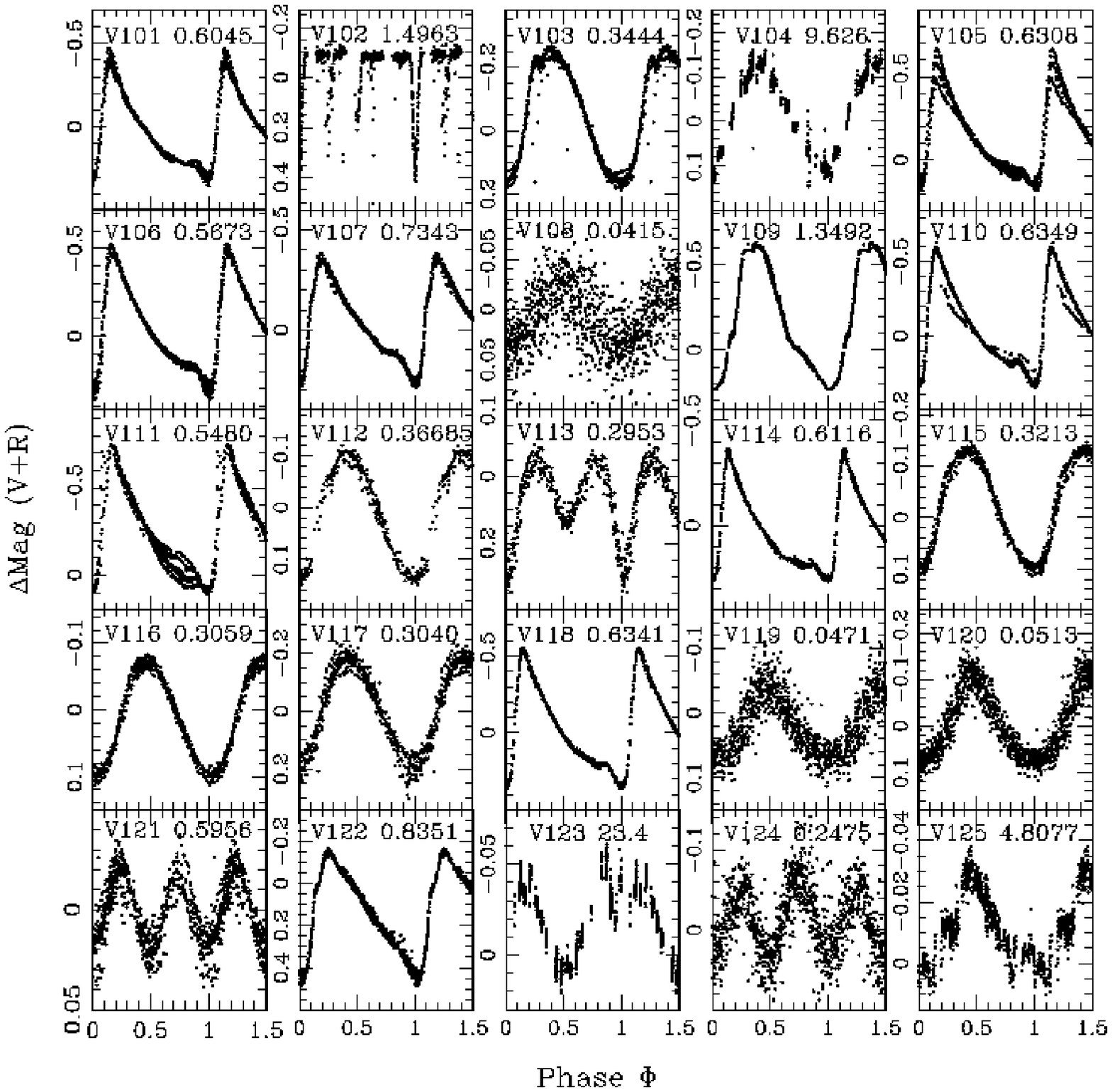}
\figcaption[figure13.eps]{Phase-wrapped differential V+R magnitude variable star lightcurves (continued).\label{varplot5}}
\end{figure}

A comparison in variable type was also made. For the 106 recovered variables, all but three have been assigned the same classification as in \citet{K2004}. These three (V42, V43 and V121) are classified as eclipsing binaries in this work, but were classed as RR Lyrae in \citet{K2004}. The time-series for these three can be seen in Fig.\space\ref{varplot2} and \ref{varplot5}. For the cases of V42 and V121, the primary and secondary eclipses have a different depth (and shape), indicating their likeliness as an binary system. V43 seemingly displays secondary variations at times of maximum brightness, as seen in other binary lightcurves (ie, V13 and V178). Two of our variables were classified as unknown in the \citet{K2004} catalog, and here are classified as an eclipsing binary (V84) and a long period variable (V104). 

\subsection{Color Magnitude Distribution}
\begin{figure}
\plotone{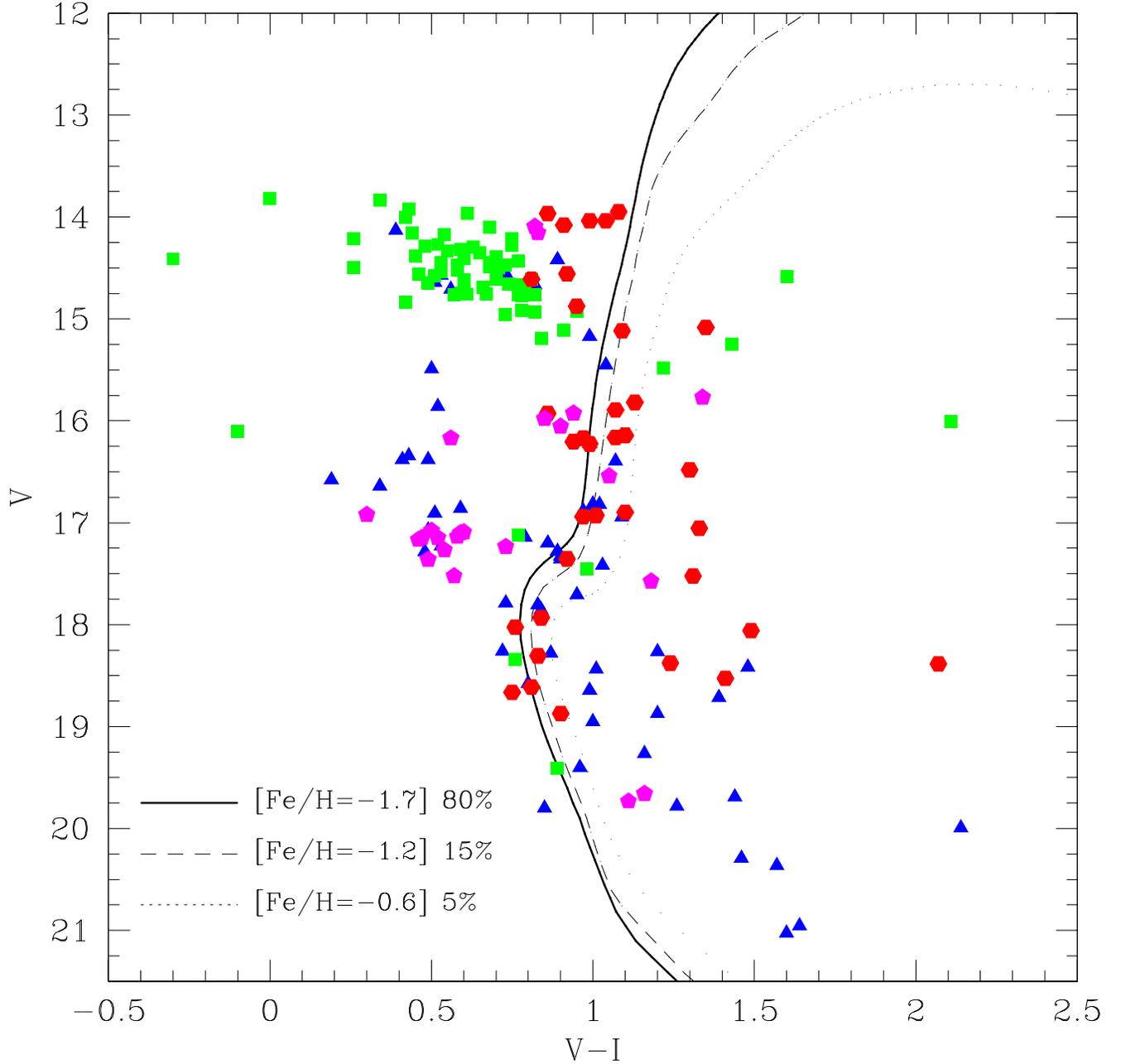}
\figcaption[figure6.eps]{The theoretical CMD for $\omega$ Cen, calculated using Y2 isochrones \citep{Y2003}, with our detected variable stars overplotted. Three isochrones are plotted, with metallicity values taken from \citep{N2004} to describe the total stellar population of the cluster. The fraction of stars that belong to each isochrone is also marked. Eclipsing binaries (EcB) are plotted as blue triangles, RR Lyrae as green squares, long period variables (LPVs) as red hexagons. The other miscellaneous variables (SX Phe stars and other pulsators) are marked as magenta pentagons. Clearly, different populations of variable stars are seen in the dataset. The EcB run from the blue straggler regime (where also most of the SX Phe stars lie) and follow the binary main sequence. Hence the majority are expected to be members of the cluster. The RR Lyrae stars are located within the Horizontal Branch instability strip and most are expected to be cluster members. The LPV stars appear to be positioned on the cluster red giant branch with some foreground variables.\label{linecmd}}
\end{figure}

Fig.\space\ref{linecmd} shows the distribution of the variable stars overlaid on the \citet{Y2003} theoretical stellar isochrones of $\omega$ Cen as determined with parameters taken from \citet{N2004}. Three isochrones were produced in total, each with the differing metallicities of the three distinct populations observed in $\omega$ Cen.

Eclipsing binaries are marked as blue triangles, RR Lyrae are plotted as green squares, the Long Period Variables are plotted as red hexagons, with the miscellaeous variables ($\delta$ Scuti's and other pulsators) marked as magenta pentagons. The eclipsing binary sample can be seen to follow the expected locations of blue straggler stars and binary main sequence members (located redward of the cluster main sequence). A few systems appear to lie on the cluster subgiant/red giant branch; if they are members, these stars are likely composed of at least one evolved component. Three systems seemingly lie on the cluster MSTO, although none are detached systems. These are hence less suitable than detached systems for determining the properties of turnoff stars in the cluster.

The RR Lyrae stars are seen to lie in the vast majority on the cluster HB instability strip, strongly implying their cluster membership. The other RR Lyrae lying off this sequence (those fainter and redder) we have classified as Galactic halo contamination (see Section\space7.2). Many of the detected Long Period Variables (LPVs) appear to be associated with the cluster red giant branch, indicating their likely nature as evolved pulsators with cluster membership. 

\subsection{Spatial Distribution}
\begin{figure}
\plotone{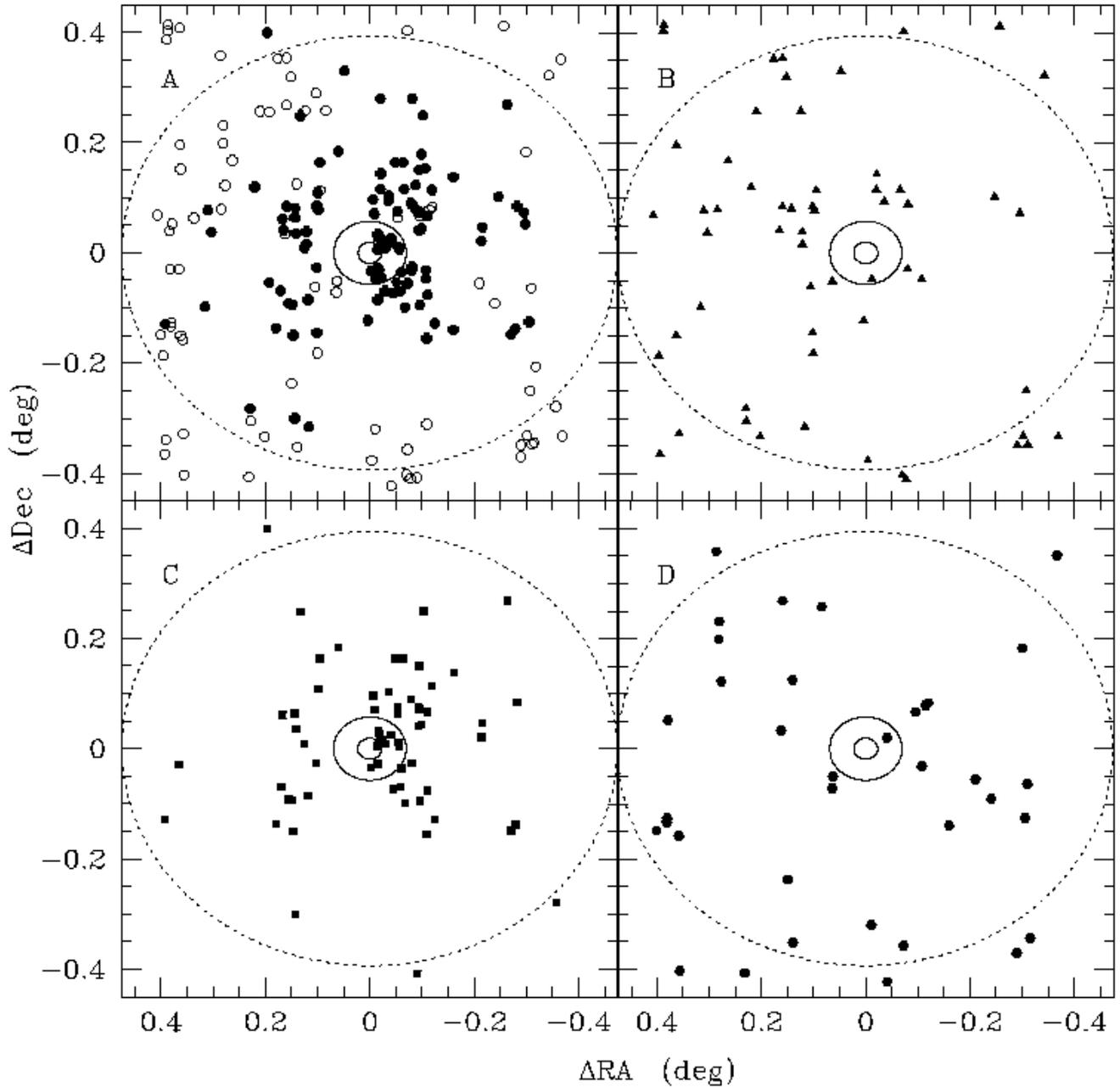}
\figcaption[figure7.eps]{The spatial distribution of the detected variable stars, measured as $\Delta$RA and $\Delta$Dec (in degrees) from the core of $\omega$ Cen. The plots have been made to the same scale as in Fig.\space\ref{clusterastrom}. The two inner ellipses correspond to the core and half mass radii of the cluster, and the large outer ellipse defines 50$\%$ of the cluster tidal radius, indicating the extent of the dataset. Panel 'A' displays all variable stars, with new discoveries marked as open circles. Panel 'B' shows the distribution of eclipsing binaries, panel 'C' the RR Lyrae stars, and panel 'D' the long period variables.\label{varsastrom}}
\end{figure}

The spatial distribution of the variable catalog, measured as the $\Delta$RA and $\Delta$Dec in degrees for each variable from the core of the cluster, is presented in Fig.\space\ref{varsastrom}. Also marked as ellipses are the core radius (inner ellipse), the cluster half-mass-radius (middle ellipse) and 50$\%$ of the cluster tidal radius (outer ellipse, the limit of the search) produced with the same scale as Fig.\space\ref{clusterastrom}. Each panel of Fig.\space\ref{varsastrom} displays the distribution of a different type of variable. 

Panel `A' shows the distribution of the total catalog: those variables recovered from the catalog of \citet{K2004} are marked as filled circles, new discoveries are hence open circles. The new discoveries are located mainly in the outer regions of the field, due to the wide field employed in our search. The apparent gap in the distribution seen at $\Delta$RA$\sim$0.05 is not real, but is due to the incompleness of the lightcurve database at this location caused by telescope drift during observations. These limits to the regions where lightcurve production was imparied can be seen as the darker shading on Fig.\space\ref{clusterastrom}. The vast majority of previously known variables that were not identified in our survey are either located in these parts of the dataset in the inner core of the cluster, a region to which this work is not sensitive. Panel `B' displays the spatial distribution of the detected EcB stars. It can be seen that a slight majority (60$\%$) are located towards the west of the cluster core. Panel `C' shows the corresponding distribution of the detected RR Lyrae stars. These stars appear to be more centrally concentrated than the other types of variables, but this is due to incompleteness of fainter stars towards the cluster core, which truncates the frequency of the fainter EcB and LPV variables. As the EcB and LPV's appear more homogeneously distributed over the cluster field, there do seem to be more EcB and LPV's than RR Lyrae in the outer regions of the cluster. Panel `D' is the distribution of the LPV stars.  

\subsection{Catalog Detection limits}
\begin{figure}
\plotone{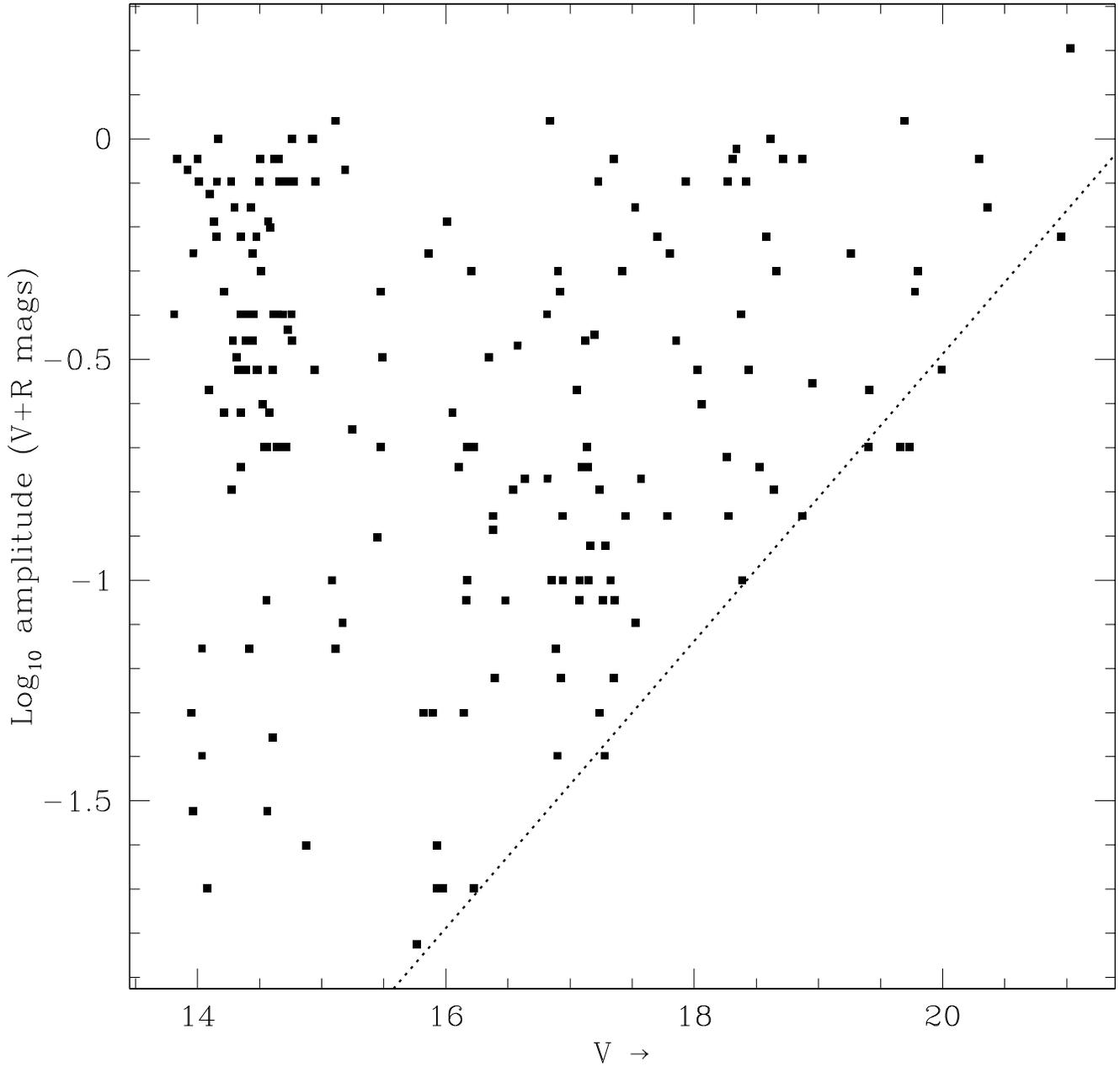}
\figcaption[figure8.eps]{The logarithm of the total amplitude of each variable star (measured with the V+R filter), plotted as a function of corresponding V magnitude. This allows insight to be made into the detection limits of the variability search. It can be seen that the catalog detection limit is 0.015 magnitudes (1.5$\%$)at V$\sim$16.0. Therefore variables with total amplitude greater than or equal to this value at this magnitude can be detected. Similarly, at V$\sim$20.0, the limit of detection is measured at $\sim$0.3 magnitudes (33$\%$), and represents the limits imposed by both the photometric uncertainty and the AoV variability detection method.\label{amplitudes}}
\end{figure}

Fig.\space\ref{amplitudes} presents the total amplitude of each detected variable star plotted as a function of its corresponding V magnitude, in order to provide information on the detection limits of the catalog. Any variable star with an amplitude greater than or equal to the position of the dotted line is very likely to be detected in the dataset by the two detection methods described previously. Hence, at V$\sim$16.0, this detection limit is about 0.015 magnitudes (1.5$\%$), while at V$\sim$20.0, the detection limit is 0.31 magnitudes (33$\%$).

\section{Preliminary Analysis of Individual Variables}
\begin{figure}
\plotone{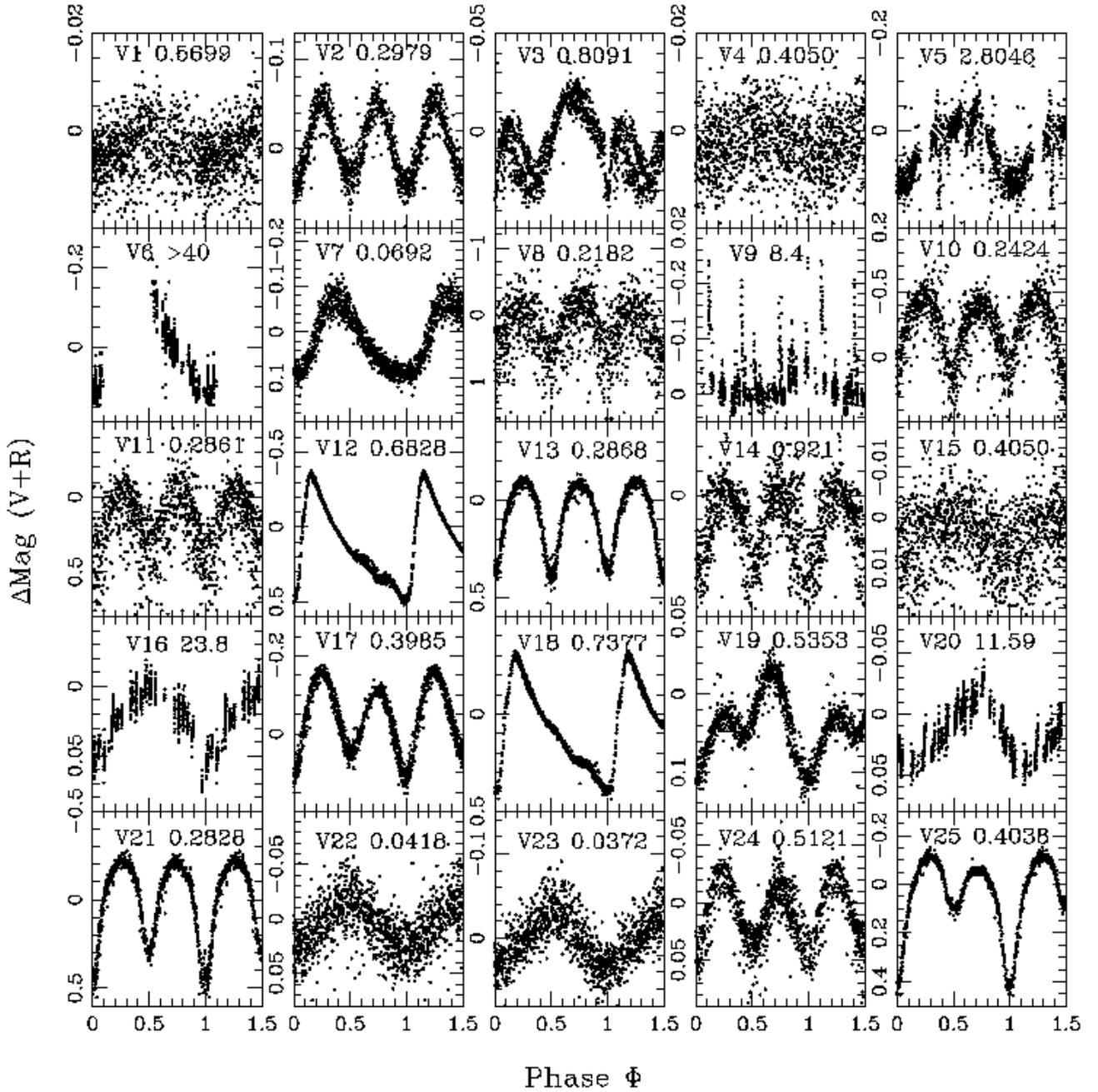}
\figcaption[figure9.eps]{The phase-wrapped differential V+R magnitude lightcurves for all variable stars detected in the dataset, plotted in order of identification. The designation of each star and the respective period in days are also plotted. \label{varplot1}}
\end{figure}

\begin{figure}
\plotone{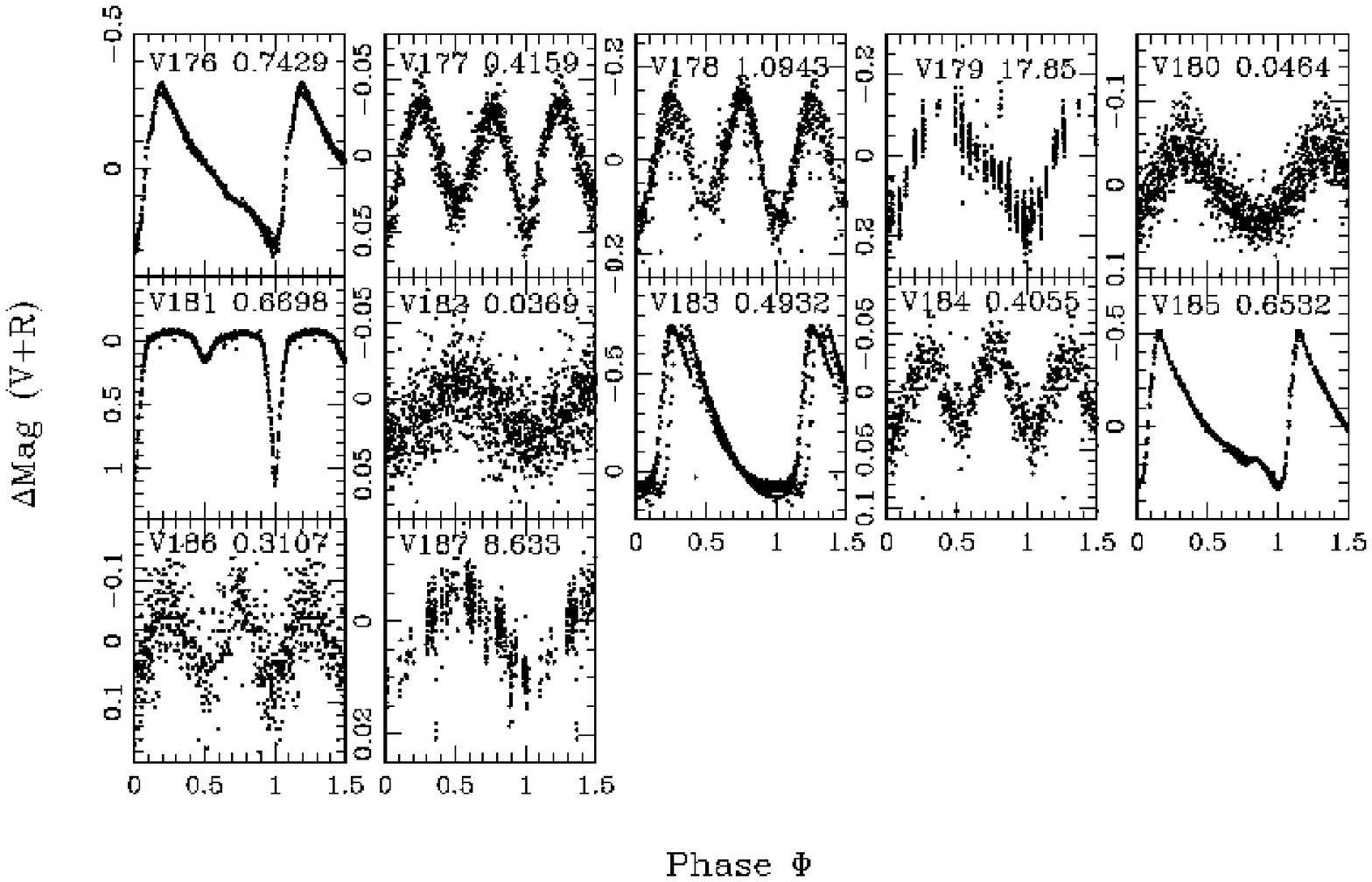}
\figcaption[figure16.eps]{Phase-wrapped differential V+R magnitude variable star lightcurves (continued).\label{varplot8}}
\end{figure}

Phase-wrapped differential V+R magnitude lightcurves of the detected variable stars are presented in Figs.\space\ref{varplot1} $-$\space\ref{varplot8} in order of discovery. Overplotted for each variable star is its identification number and determined period in days. The catalog as a whole is tabulated in Table\space2. Those variables marked with `$-$' in the last column of Table\space2 are new discoveries.

\subsection{Eclipsing Binaries}
For the 58 EcB stars, it is clear that examples of contact (ie, V2), semi-detached (ie, V37) and detached (ie, V59) configurations are all present in the $\omega$ Cen field. These are classified considering the presence (or lack thereof) of features in the time-series indicating tidal distortion of the stellar components, the derived period, and the presence of any datapoints between individual eclipses, indicating physical separation of the stars. The majority (75$\%$) are contact W-UMa-type systems, with apparent blue stragglers, binary main-sequence members and foreground variables identified from their locations on the cluster CMD (see Fig.\space\ref{linecmd}). A further sample lie on the cluster Red Giant Branch (RGB), indicating that if they are cluster members they likely contain at least one evolved component.

\begin{figure}
\plotone{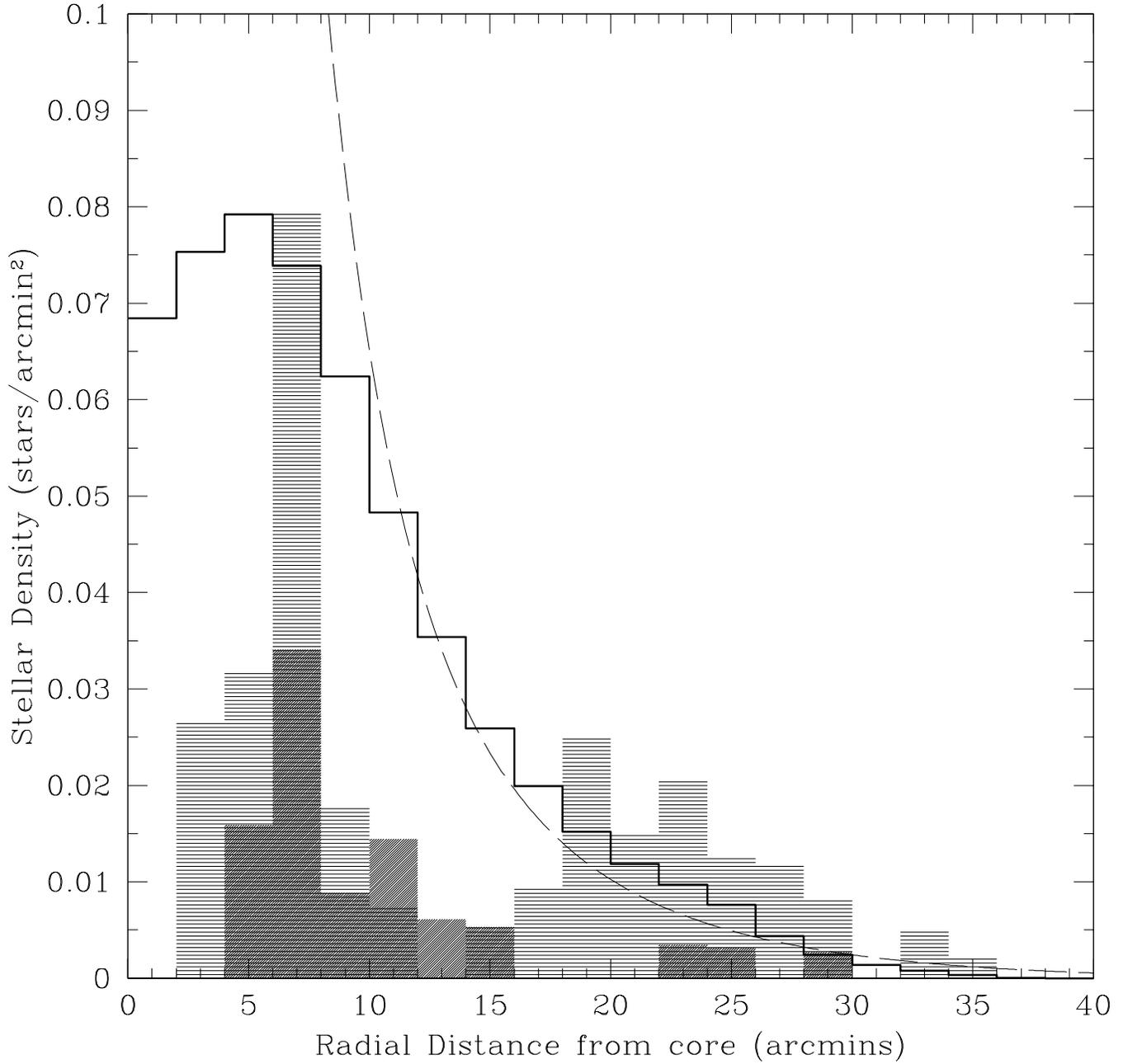}
\figcaption[figure17.eps]{The radial distribution of eclipsing binary stars, plotted as binary density as a function of projected distance in arcminutes from the cluster core. The light histogram corresponds to the contact binaries (P$\le$1d) and the darker shading the detached binaries (P$>$1d). There is no difference between the radial distribution of these different types of binary in the $\omega$ Cen field as indicated by a KS test. Also plotted is the total stellar population (open histogram) and the theoretical King profile using the parameters of \citet{Harris96}. Our recovered stellar densities are complete to R$\sim$10$'$. There appears to be a lack of binaries with radial distance 8$'$$\rightarrow$15$'$ in the cluster, perhaps indicating two separate populations.\label{radhist}}
\end{figure}

Fig.\space\ref{radhist} presents the radial distribution of EcB, measured as the distance in arcminutes from the position of the core of the cluster. The light grey shaded histogram represents the radial distribution of the contact and semi-contact EcB (with periods $\le$1 day), while the dark shaded histogram denotes the detached binaries. It can be seen that there is a marginal difference in the distribution of these two different types of binary systems; the results of a KS-test show that there is a 63$\%$ probability that they have the same distribution. In contrast, 47 Tucanae shows a clear segregation with contact binaries being preferentially located towards the cluster core with high significance \citep{W2004}. Also overplotted is the normalised total stellar distribution of the cluster (open histogram) and the theoretical King Profile \citep{K1962}, as derived using the cluster parameters taken from \citet{Harris96}. The total stellar distribution is seen to have a 100$\%$ completeness limit from 10$'$ radius outwards. 

The binary radial distribution shows an apparent gap in the population in the range 8$'$$\rightarrow$15$'$ from the cluster core. Other variable star types do not show this gap in their distributions, and so the effect is not thought to be attributed to variable star recoverability limits. From analysis via a KS test, there is only a 10$\%$ chance that the binaries are distributed in the same way as the other types of variable identified. This may be evidence for two populations of binary systems in the cluster and cannot be attributed to systematic dataset completeness limits. By compensating for the gaps in the time-series spatial coverage caused by telescope offsets during the observing run (hence regions where no time-series were produced), this binary distribution is enhanced rather than diminished.

This gap in binaries cannot be attributed to mass segregation as the cluster does not show any evidence of this \citep{A1997,DSR2005,Feretal2006}. It is possible that this distribution is related to the early life of the cluster as the nucleus of a dwarf galaxy \citep{BF2003,IM2004,BN2005}, which produced a base level of primordial binaries from gas-feeding, which we see as the outer population. Subsequent globular cluster processes (tidal and 3-body) would account for the increase in binarity towards the core. As the cluster relaxation time increases with radial distance due to longer interaction timescales, it is possible that the outer population is composed of the original primordial binaries (K.C.Freeman, private communication). The gap in the EcB distribution could indicate the boundary between these two effects, but it remains unclear with current data.

\subsubsection{Peculiar Eclipsing Binaries}
A number of binaries display interesting features in their lightcurves. Two systems have magnitudes and colors that place them on the cluster Horizontal Branch Instability Strip (hence in the same location as the RR Lyrae stars): V3 and V166. These could be systems that contain one pulsating component orbited by a secondary star (if members), or alternatively (perhaps more likely) they could be composed of one heavily distorted component with a small high-mass companion of foreground Galactic halo membership. 

The first system, V3, has a short orbital period of 0.81 days and displays continuous sinusoidal variability with an eclipse of $\sim$0.04 magnitudes depth. The eclipsing companion is strongly distorting the primary star. There seem to be two separate sequences of sinusoidal variability with the same period as the binary companion and an indication of an extended region of eclipse immediately before and after the main eclipse, perhaps indicating the presence of an accretion disk. 

The second system of this same type, V166, is similar, displaying the same strong continuous sinusoidal variability with an eclipse (of $\sim$0.16 magnitude depth), but with a longer orbital period of 2.06 days and much higher amplitude of variability. This star also shows the same two sequences of sinusoid, one sharp-edged and another more rounded with the same period as the binary companion. Both systems are worthy of future spectroscopic follow-up to determine their true nature. 

A further binary system of note, V39, displays a single eclipse in the dataset with a total eclipse duration of $\sim$3 days. Our data do not allow conclusions on the central shape of the eclipse. The color and magnitude of this system is consistent with a cluster member located high on the red giant branch, hence composed of at least one evolved giant star, consistent with an eclipse of long duration.

\begin{figure}
\plotone{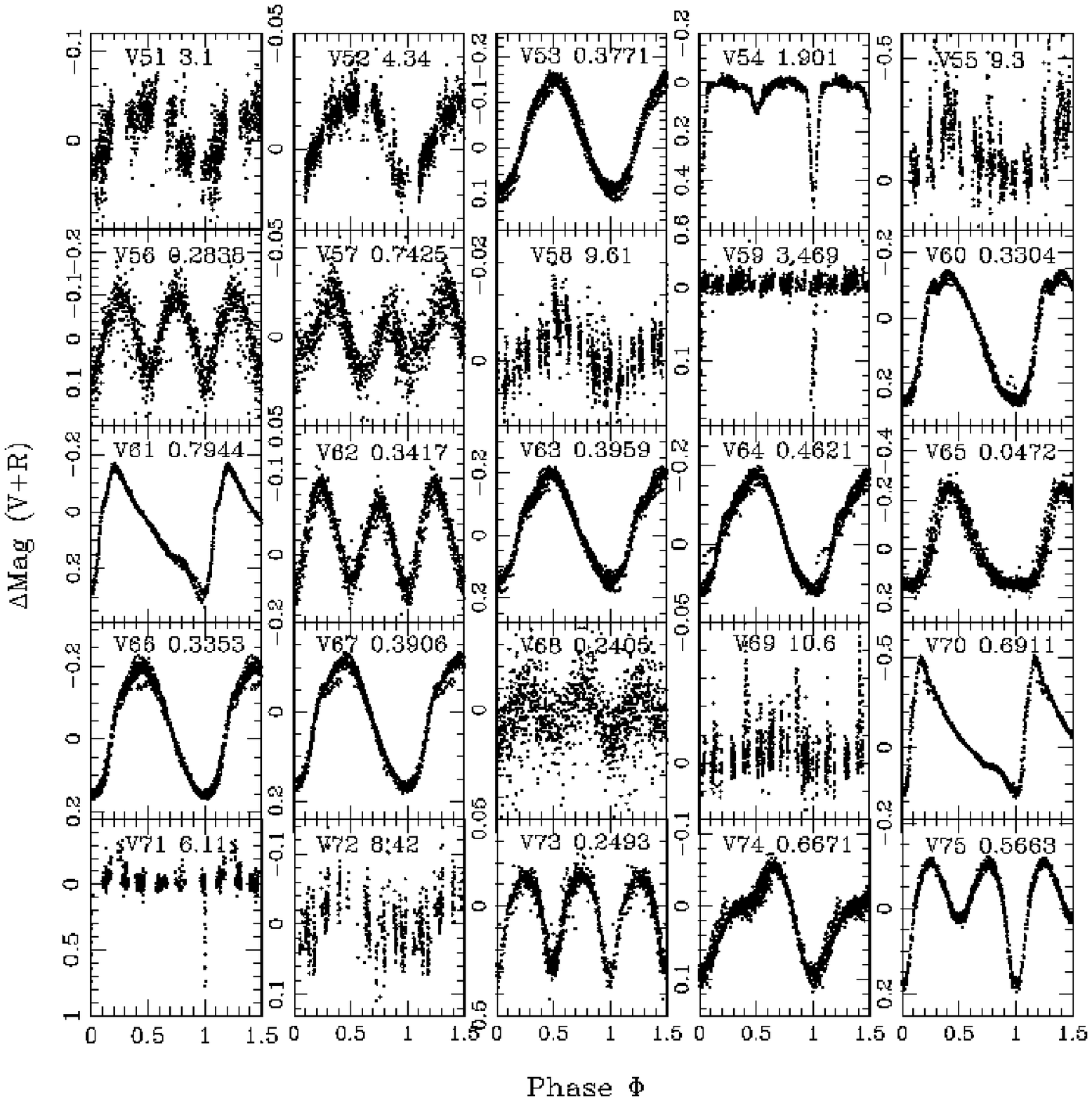}
\figcaption[figure11.eps]{Phase-wrapped differential V+R magnitude variable star lightcurves (continued).\label{varplot3}}
\end{figure}

\begin{figure}
\plotone{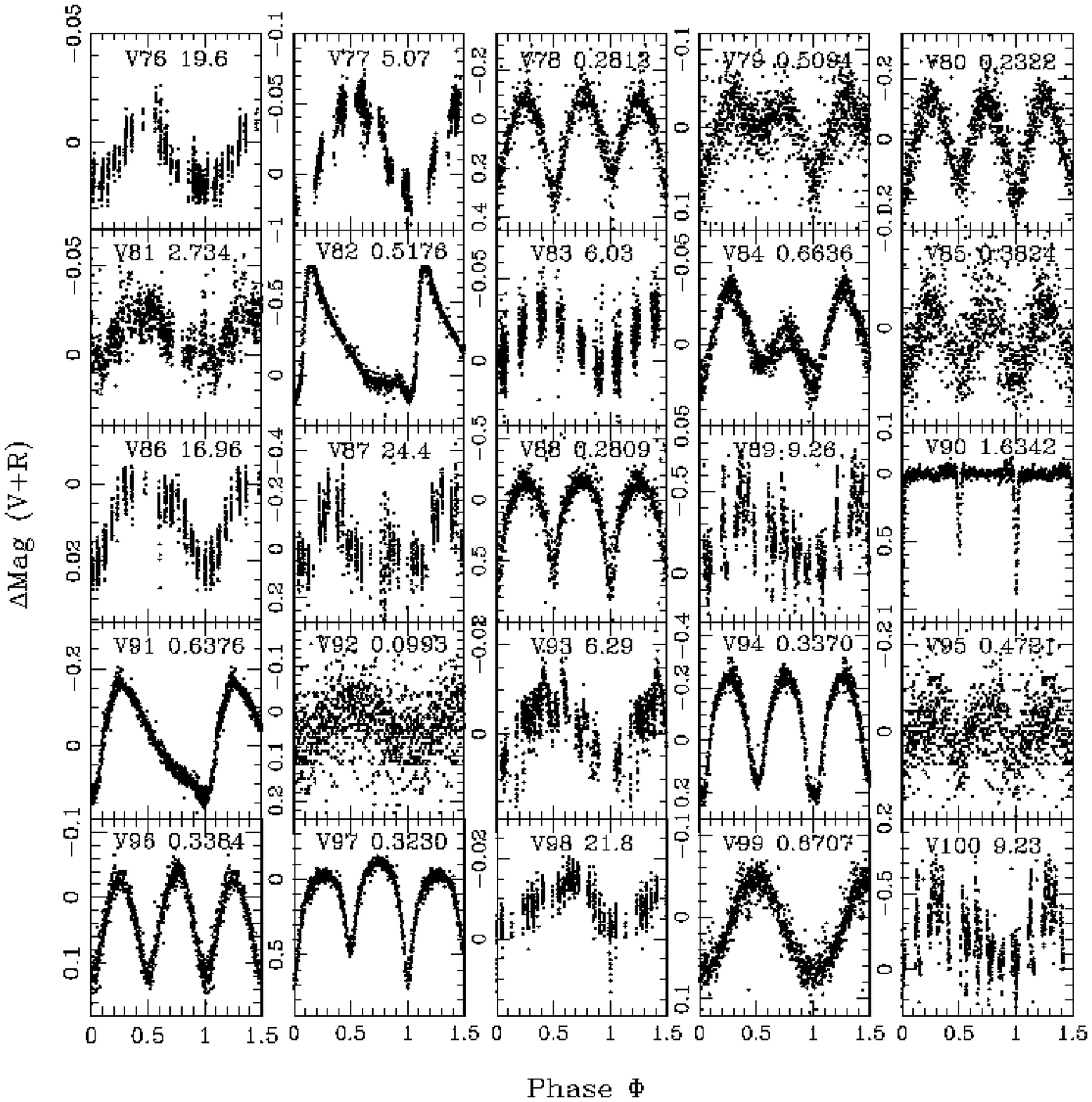}
\figcaption[figure12.eps]{Phase-wrapped differential V+R magnitude variable star lightcurves (continued).\label{varplot4}}
\end{figure}

\begin{figure}
\plotone{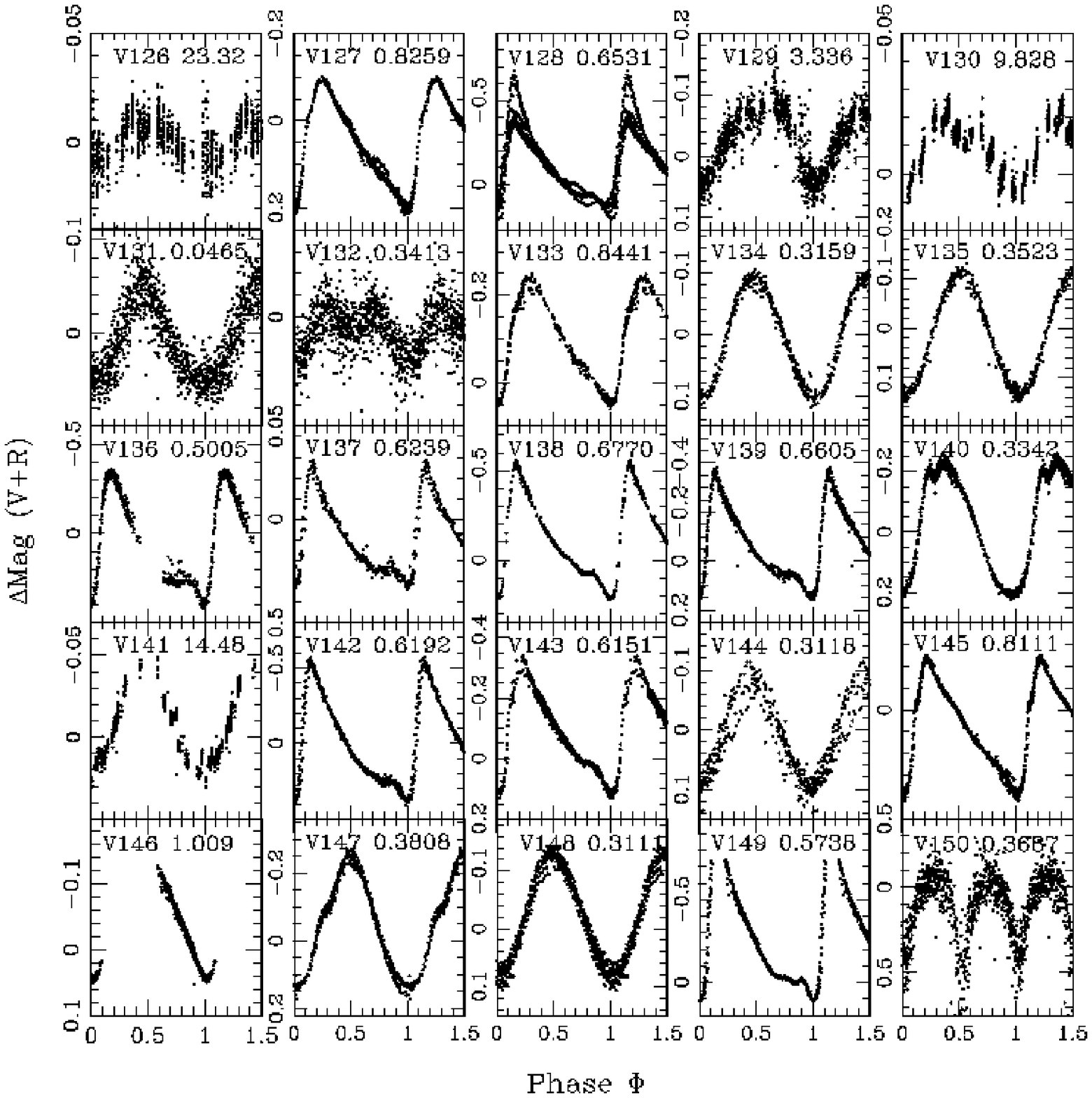}
\figcaption[figure14.eps]{Phase-wrapped differential V+R magnitude variable star lightcurves (continued).\label{varplot6}}
\end{figure}

\begin{figure}
\plotone{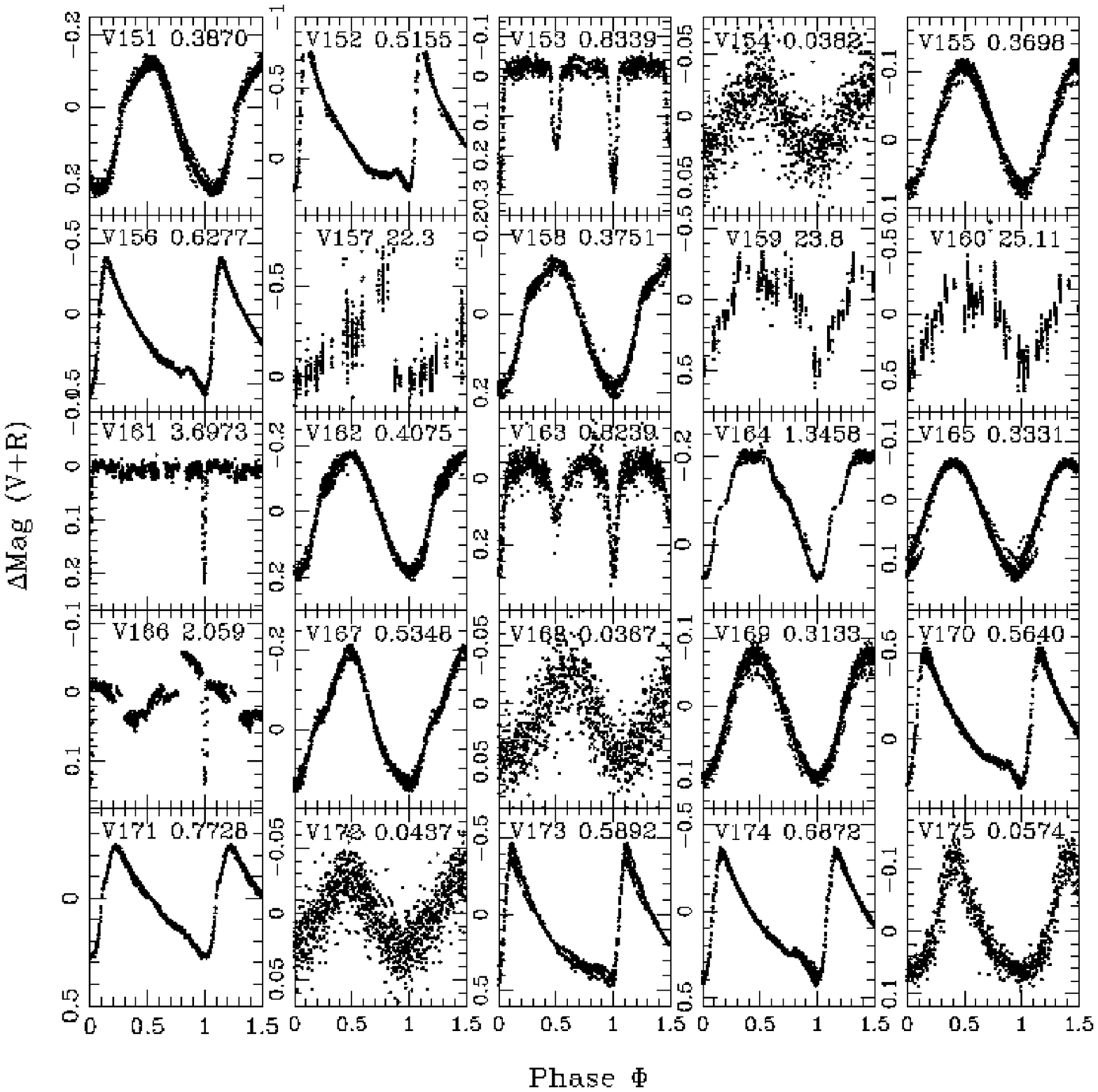}
\figcaption[figure15.eps]{Phase-wrapped differential V+R magnitude variable star lightcurves (continued).\label{varplot7}}
\end{figure}

The derived period of this binary system, 34.8 days, has been determined by the appearance of secondary variations caused by tidal distortion of the primary star (even though only one eclipse was seen). The location of any secondary eclipse has been given away by the small-scale drop in brightness seen on these secondary variations. In this same way, other single eclipsers in the dataset have had their likely periods determined (ie, V71 and V161). These secondary-determined periods are denoted in Table\space2 with a $\ast$, and the sinusoidal variability for these stars can be seen in their lightcurves in Figs.\space\ref{varplot1}$-$\ref{varplot8}.

\subsubsection{Low-Mass Eclipsing Binaries}
Of the contact eclipsing binary sample, seven systems have orbital periods $\le$0.25 days; V8, V10, V30, V68, V73, V80, and V124. Due to these very short periods, it is expected that these systems are composed of low-mass components, very likely late K to M dwarf stars. All of these systems have magnitudes and colors that overlay on the $\omega$ Cen binary main sequence, and hence most are likely members of $\omega$ Cen itself. Of this sample, V8, V10, V30 and V80 have V magnitudes in the range 19$\rightarrow$21. The other stars are located approximately at the cluster turnoff.

Three detached binaries have orbital periods $\le$1.6 days (V90, V102 and V153), with a further system with P$=$2.46d (V41). These have been classified as detached systems due to the lack of observable secondary variations caused by the tidal distortion of the components occuring outside of the main eclipses and by the differing eclipse depths observed. These systems must be composed of low-mass components (perhaps both in the M-dwarf regime) for them to display such short periods without the effects of tidal distortion on the lightcurves. The identification (and confirmation) of M-dwarf detached binaries is of great importance in the production of stellar evolutionary models since they allow determination of masses and radii towards the low end of the mass function for comparison to theoretical predictions.

Of particular interest is V153, a detached binary with an orbital period of 0.83 days. The color and magnitude of this star place it blueward of the blue-straggler branch of the CMD, and hence in an unusual place for a detached binary member of the cluster (which should lie on one of the main CMD sequences). In fact, the V-I value of 0.19 is unusually blue for a detached binary of such short period, which would ordinarily be composed of red components. From our observations of the period and color, considering the distinct observed eclipses (with somewhat flat bottoms) and lack of any observable tidal distortion, we conclude that this binary is very likely a foreground system, with an uncertain composition, perhaps containing a pair of white dwarf stars. Such a system would be a prime candidate for future spectroscopic observations to determine its true nature.

\subsection{RR Lyrae}
A total of 69 RR Lyrae stars were identified in the dataset, the vast majority of which are likely members of the cluster, due to their firm location on the cluster Horizontal Branch. Of the total sample, 59$\%$ are examples of RR Lyrae type `AB' (fundamental mode pulsators) and 41$\%$ are examples of the shorter period RR Lyrae type `C' (first harmonic mode pulsators) based on the shape of their phase-wrapped lightcurves and periods. Type `C' stars are typically less common than Type `AB' \citep{Vivas01}. None of the rare longer period RR Lyrae AHB1 stars \citep{Sand94} were seen in our data. Twenty-six of our RR Lyrae are new discoveries.

\begin{figure}
\plotone{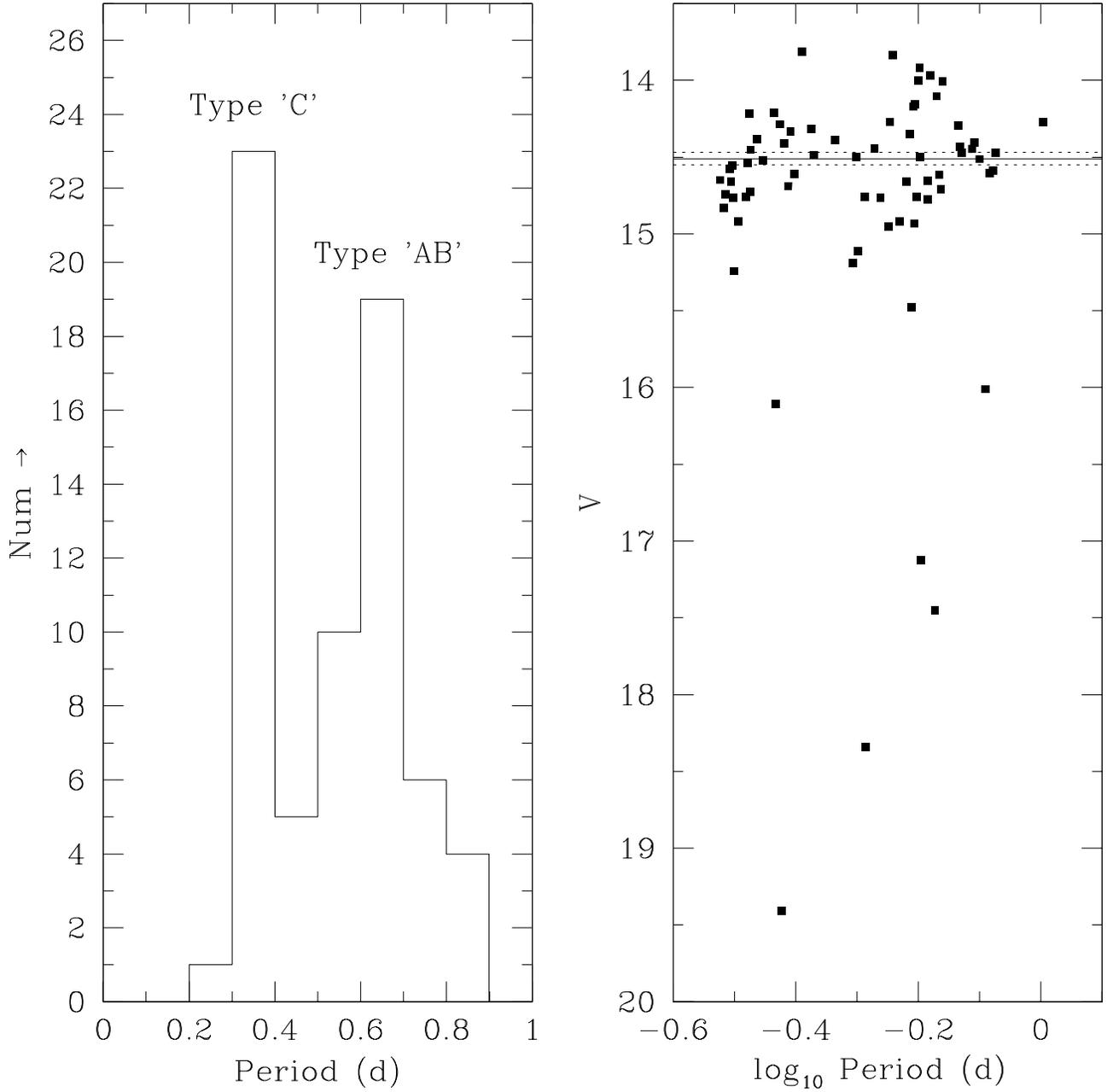}
\figcaption[figure18.eps]{The period distribution of the detected RR Lyrae stars (left), and the corresponding period-magnitude diagram (right). Two main populations of RR Lyrae can be seen: the longer period Type `AB' RR Lyrae and the shorter period Type `C'. The period-magnitude diagram shows that all the RR Lyrae are clustered around V$=$14.51$\pm$0.04, strongly indicative of their membership in $\omega$ Cen. The shorter period Type `C' RR Lyrae show somewhat less scatter around their mean than their longer period Type `AB` counterparts. This is due to the `C' RR Lyrae having smaller amplitudes, thus magnitudes of these stars at random phases are closer to their actual average magnitude. A number of RR Lyrae are seen running from V$\sim$16.0$\rightarrow$$\sim$19.5 and are attributed to background contamination from the Galactic Disk. No RR Lyrae are seen brighter than V$\sim$13.8 as our time-series data are saturated there (see Fig.\space\ref{rmsplot}).\label{rrlyrplots}}
\end{figure}

\begin{figure}
\plotone{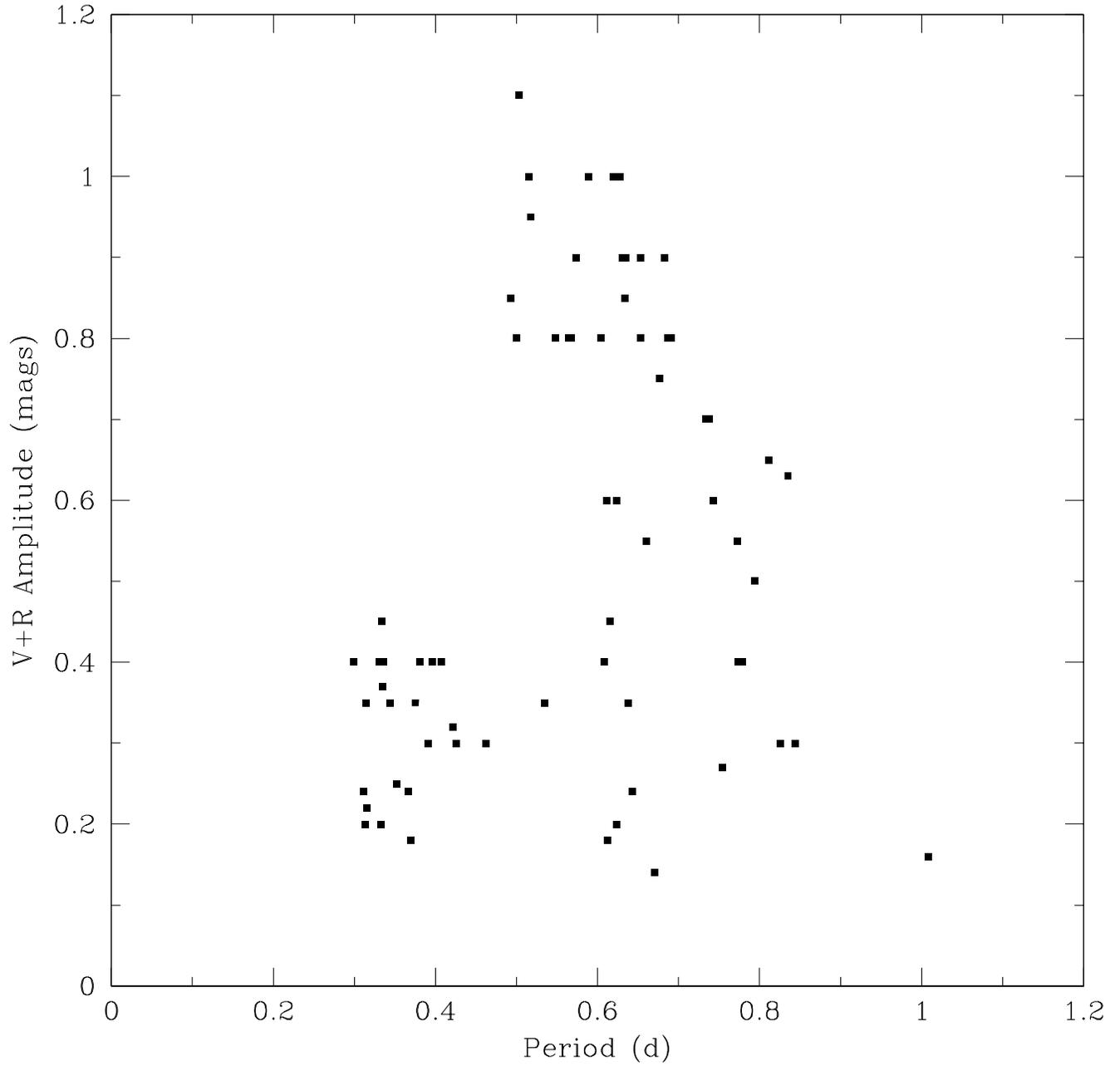}
\figcaption[figure19.eps]{The period versus V+R amplitude plot for our sample of RR Lyrae stars.\label{rrlyramps}}
\end{figure}

Fig.\space\ref{rrlyrplots} shows the period distribution (left panel) and the period-luminosity diagram (right panel) for our RR Lyrae sample. The period distribution displays two distinct populations, one peaking at pulsation period $\sim$0.35 days and the other at period $\sim$0.65 days. These constitute the RR Lyrae type `C' and type `AB', respectively. Fig.\space\ref{rrlyramps} shows the period versus V+R amplitude plot for our sample of RR Lyrae stars.

The period-luminosity diagram shows that the majority of RR Lyrae (of both types) are clustered around V$\sim$14.5 and are hence likely $\omega$ Cen members. A small number of the sample are significantly fainter, running from V$\sim$15.0$\rightarrow$19.5. The outliers are attributed to background contamination by the Galactic Halo. Our photometry becomes saturated at V$\sim$14.0 (as seen in Fig.\space\ref{rmsplot}) and hence there are no foreground RR Lyrae detected in the dataset with brighter V magnitudes than this. 

To determine memberships based on the magnitude distribution, an application of KMM mixture modelling \citep{Ashman94} was applied, which assumes gaussian distributions with differing dispersions. As a first check, the faintest six RR Lyrae were removed using 3-sigma clipping, producing a mean V magnitude of 14.50$\pm$0.33 for the remaining 63 stars. In comparison, KMM modelling was applied to the total sample with two gaussians, which produced two samples, one with the 62 brightest RR Lyrae and another with the remaining 7 fainter stars. The result is that everything brighter than and including V$=$15.245 (62 stars) is considered a cluster member based on its brightness. The mean of these 62 (for both AB and C type RR Lyrae) is 14.51$\pm$0.04, with the error being the standard deviation of the mean, seen overplotted in the right panel of Fig.\space\ref{rrlyrplots}.

From our total RR Lyrae sample, it is possible to calculate the distance modulus of the cluster. It is well known that the absolute magnitude of RR Lyrae stars is dependent on metallicity \citep{Sand1981a,Sand1981b}, and from studies of RR Lyrae and horizontal branch stars in the Milky Way, the LMC and M31 clusters this relation has been recently adopted with a slope in the range 0.20-0.23 mag/dex \citep{Cl2003,Ca2003,G2004}. \citet{R2005} have adopted a relation of the form $M\rm{_{V}}$$=$0.20$\pm$0.09[Fe/H]$+$0.81$\pm$0.13, which is also used in this work.

\begin{figure}
\plotone{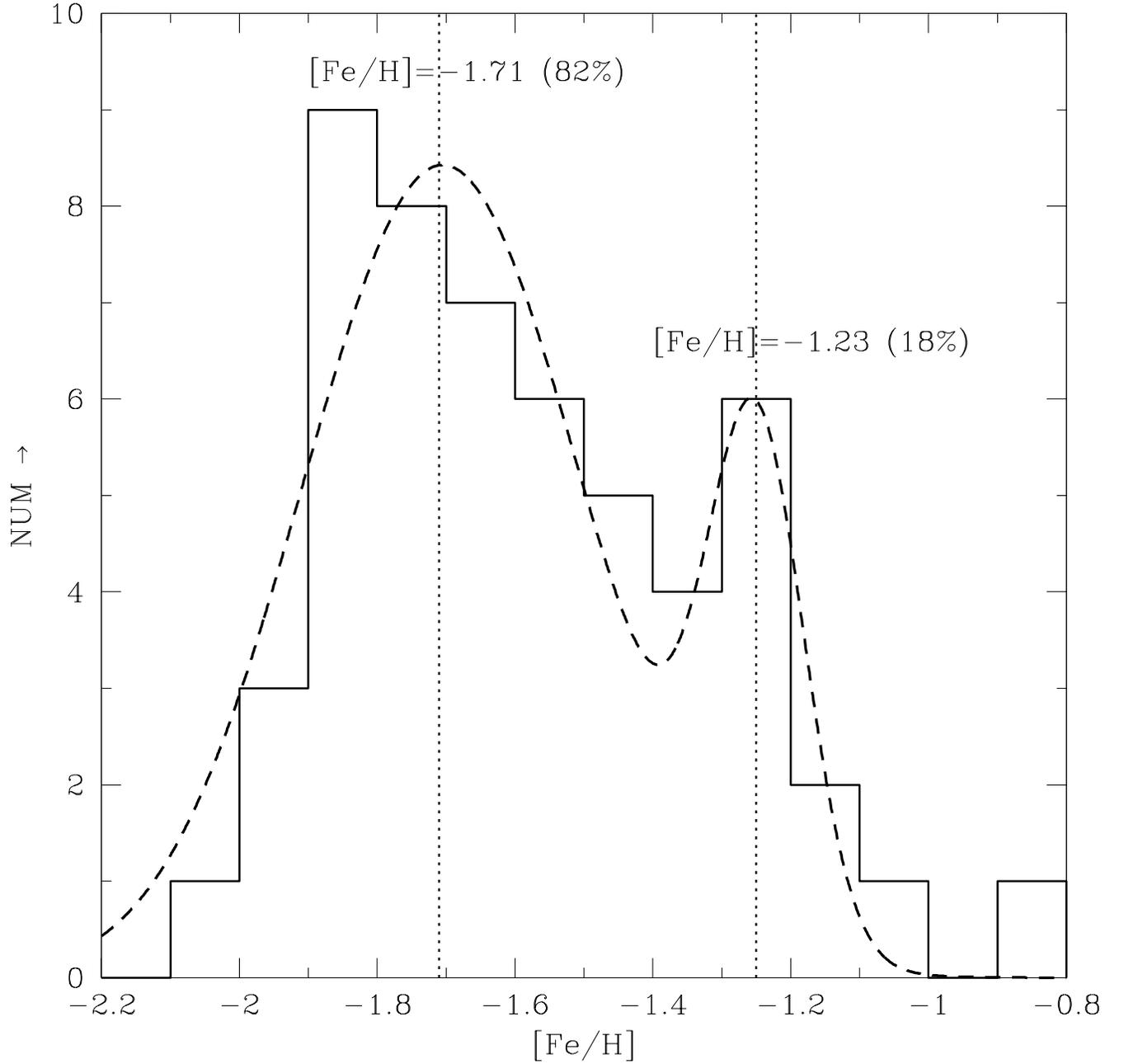}
\figcaption[figure20.eps]{The distribution of [Fe/H] for 53 of our RR Lyrae stars which could be crossidentified with those of \citet{Rey2000} (open histogram). Two Gaussian distributions have been fitted and their peak [Fe/H] and relative fractions determined (overplotted), with a dispersion due to the natural dispersion of the sample and the error in the \citet{Rey2000} metallicity determinations. This [Fe/H] distribution is assumed to apply to our total RR Lyrae sample and is used to determine the cluster distance modulus.\label{RRLyrmets}}
\end{figure}

Several of our RR Lyrae sample could be crossidentified with the sample of \citet{Rey2000}, which have metallicity measurements associated with them. Fig.\space\ref{RRLyrmets} shows the [Fe/H] distribution for the 53 of our RR Lyrae sample, which were crossidentified with these \citet{Rey2000} metallicity estimates. Two Gaussians were fitted via least squares to describe this distribution. The resulting Gaussians are overplotted in Fig.\space\ref{RRLyrmets}, with their peak [Fe/H] and relative fractions noted. From the area under the gaussian fits, 82$\%$ of the sample belong to a population with [Fe/H]$=$-1.71 with the remaining 18$\%$ belonging to a [Fe/H]$=$-1.25 population. The dispersions of these two Gaussians are measured as 0.20 and 0.07 dex respectively, being composed of the natural dispersion of the sample and the uncertainty in the [Fe/H] measurements as presented by \citet{Rey2000}. We assume that the remaining 16 RR Lyrae in our sample with unknown [Fe/H] follow these same relative distributions.

The absolute magnitude (M) of the most metal poor population is 0.47$\pm$0.28 as determined via the \citet{R2005} relationship, with M$=$0.56$\pm$0.24 for the second more metal rich group. By considering the relative fractions of the sample in each population, the weighted mean absolute magnitude estimated for our total RR Lyrae sample is 0.48$\pm$0.27. 

The mean V apparent magnitude for the cluster RR Lyrae sample (of all metallicities) is 14.51$\pm$0.04 with the error being the standard deviation of the mean. This leads to an apparent distance modulus of 14.04$\pm$0.27 (6.4$\pm$0.7kpc). This compares well to the \citet{Harris96} value of 13.97 and the \citet{Kal2002} estimate of 14.09$\pm$0.04 as derived from an eclipsing binary star. Our apparent value is around 0.3 magnitudes brighter than that of \citet{VDV2005}, who find a best-fit dynamical distance modulus of 13.75$\pm$0.13. The cluster has a non-zero reddening, and if we apply the E(B-V)$=$0.12 value of \citet{Harris96}, we arrive at a reddening-corrected distance modulus of 13.68$\pm$0.27 (5.4$\pm$0.7Kpc). This compares well to the reddening-corrected estimate of 13.70$\pm$0.06 from \citep{DP2006}.

A small number of the RR Lyrae appear to display small-scale secondary variations in the general shape of the lightcurve and are examples of Blahzko RR Lyrae \citep{Bla07}. Examples of this effect are seen in V48, V105, V110, V111, V128, V136, V173 and V183. Quite significant variations are seen particularly in V111, V128 and V183. Such variations are a little understood feature of some RR Lyrae, and are generally thought to be caused by strong photospheric magnetic fields \citep{Cous83} and/or rotation \citep{Dziem99}. The effect is dependent on the metallicity of the star \citep{Alcock03} and is found preferentially in association with type `AB' RR Lyrae. We note, however, that V183 cannot be easily phase-wrapped and seems to be undergoing a period change. Due to their sinusoidal lightcurves, their faintness compared to the cluster HB and small amplitude, it is possible that V53, V99 and V155 are eclipsing binary stars, and are classified as either RR Lyrae or EcB in Table.\space1. As only a small fraction of the lightcurve for V146 was observed, the derived period in Table\space1 is uncertain, but due to it's location on the cluster CMD it has been classified here as an RR Lyrae. Further photometry is needed to fully determine it's type.

\subsection{Long Period Variables and Miscellaneous Variables}
In this work, an LPV is defined as having a pulsation-like lightcurve with a periodicity $\ge$2 days. Compared to some types of variables (ie, Mira-type stars), the LPVs found in this dataset have very short periods. The length of our observing window (25 nights) and saturation limit precludes the detection of variable stars with excessively long pulsation periods ($\sim$years), which would exist in the bright AGB region of the cluster population. Despite this, a rich sample of LPVs with relatively short pulsation periods have been detected in the dataset.

From their locations in Fig.\space\ref{linecmd}, approximately 75$\%$ of the detected LPVs are expected to be members of the cluster. A small number ($\sim$6) seem to lie on the cluster main sequence. The rest seem too red and faint to be consistent with the cluster stellar populations. Of particular interest are V98 (P $=$21.8 days) and V125 (P$=$4.8 days). The lightcurves for these two stars appear to show eclipses superimposed on the pulsation, which we are able to discern due to the high temporal resolution of the data. Both are likely cluster members.

The miscellaneous variables are comprised of 15 SX Phoenicis ($\delta$ Scuti) stars and nine other pulsators of various types. The SX Phe are, in the majority, in the cluster blue-straggler region and hence these are regarded as likely cluster members. All have very short periods, typical of the class, with the shortest being $\sim$0.038 days. Two other pulsators, V164 and V109 have been classified as Type-II Cepheids \citep{Nem1994}. V19 could be a binary system or a pulsator. Finally, we classify V1, V4, V15 and V50 and V74 as regular short period pulsators of uncertain type.

\section{Summary and Conclusions}
We have presented the V+R lightcurves and a preliminary analysis of 187 variable stars detected in a 52$'$$\times$52$'$ field centered on the globular cluster $\omega$ Cen, as part of an extensive search for transiting `Hot Jupiter' planets in the cluster, which will be presented elsewhere. Of the variables presented here, 81 are new discoveries. The catalog in total comprises 58 eclipsing binaries, including contact, semi-detached and detached configurations; 69 RR Lyrae stars, 36 Long Period Variables, and 24 miscellaneous variables including 15 SX Phe stars and two TypeII Cepheids. From their locations on the cluster Color Magnitude Diagram, the majority of our detected variables are consistent with having cluster memberships.

The eclipsing binary radial distribution displays an apparent lack of binaries (of both contact and detached types) in the  8$'$$\rightarrow$15$'$ range, perhaps indicating the presence of two separate binary populations. This trend cannot be attributed to completeness limitations in the dataset. The origin of the spatial gap in the binaries is unclear. 

Of the total binary sample, four detached systems have short orbital periods. When combined with their observed V-I colors, three are consistent with being composed of M-dwarf stars, important for comparisons with stellar evolution models. One further detached system has an orbital period of only 0.8 days and, given its blueness, may be composed of white dwarf stars. Another eclipsing system has a single observed eclipse with a duration of $\sim$3 days, due to the presence of one evolved component. We deduce a reddening corrected distance modulus of 13.68$\pm$0.27 for the cluster, based on the RR Lyrae in our sample.

We also present an extensive V and I photometric database (with astrometry better than 0.25$''$) for 203,892 stars in our 0.75 deg$^{2}$ field of view centered on the cluster, and V+R lightcurves spanning 25 nights for 109,726 stars (14.0$\leqslant$V$\leqslant$22.0) for both the cluster and the field. 

\section*{Acknowledgments}
The authors would like to thank the following people for their contributions in the production of this work: Omer Tamuz for many discussions on the merits of the AoV detection method; Laura Stanford for help with the production of the theoretical isochrones; John Norris and Ken Freeman for discussions on the binary radial distribution; and Cristina Afonso for helpful information about KS tests. We also thank Gisella Clementini for acting as a very thorough and helpful referee.

\clearpage
\clearpage
\clearpage
\clearpage
\clearpage

\clearpage

\clearpage
\clearpage
\clearpage
\clearpage
\clearpage
\clearpage
\clearpage
\clearpage
\clearpage
\clearpage
\clearpage
\clearpage
\clearpage
\clearpage
\clearpage

\begin{center} 
\footnotesize
\begin{tabular}{lll} \hline
\noalign{\medskip}
$CCD$ & $\it{RA}(J2000.0)$ & $\it{DEC}(J2000.0)$ \\
& h\hspace{2mm}m\hspace{2mm}s & $^{\circ}$\hspace{5mm}$'$\hspace{3mm}$''$ \\
\noalign{\medskip}
\hline
\noalign{\medskip}
1 & 13:28:05 & $-$47:08:58 \\
2 & 13:28:05 & $-$47:21:52 \\
3 & 13:28:05 & $-$47:34:44 \\
4 & 13:28:05 & $-$47:47:43 \\
5 & 13:25:34 & $-$47:48:12 \\
6 & 13:25:34 & $-$47:35:06 \\
7 & 13:25:34 & $-$47:22:07 \\
8 & 13:25:34 & $-$47:08:57 \\
\noalign{\medskip}
\hline\hline
\end{tabular}  
\end{center}
\vspace*{3mm}
\normalsize{Table 1: Equatorial coordinates (J2000.0) for the centres of the eig
ht WFI CCDs.}

\clearpage


\begin{center}
\footnotesize
\begin{tabular}{lllllllll} \hline\hline
\noalign{\medskip}
$ID$ & $Type$ & $Period (d)$ & $RA (J2000.0)$ & $DEC (J2000.0)$ & $V$ & $V-I$ &
$Prev Name$ \\
&&& h\hspace{2mm}m\hspace{2mm}s & $^{\circ}$\hspace{4mm}$'$\hspace{4mm}$''$ &&\\
\noalign{\medskip}
\hline
\noalign{\medskip}
 V1  & Puls      & 0.569 & 13:29:03.20 & -47:05:22.02 & 15.77 & 1.34 & 0.02 & - \\
 V2  & EcB       & 0.298 & 13:29:02.31 & -47:04:25.78 & 16.94 & 1.09 & 0.13 &  - \\
 V3  & EcB       & 0.809 & 13:29:02.16 & -47:03:45.46 & 14.42 & 0.89 & 0.07 & - \\
 V4  & Puls      & 0.405 & 13:28:53.93 & -47:04:08.63 & 15.98 & 0.85 & 0.02 & - \\
 V5  & LPV       & 2.805  & 13:28:26.68 & -47:07:08.44 & 18.06 & 1.49 & 0.25 & - \\
 V6  & LPV   & $>$40  & 13:28:24.90 & -47:14:47.07 & 14.94 & 2.80 & $\ge$0.35 & - \\
 V7  & SX Phe & 0.069 & 13:27:53.76 & -47:13:18.45 & 16.17 & 0.56 & 0.2 & - \\
 V8  & EcB       & 0.218 & 13:27:48.17 & -47:07:27.80 & 21.03 & 1.60 & 1.4 & - \\
 V9  & LPV       & 8.4     & 13:27:42.15 & -47:12:31.77 & 16.94 & 0.97 & 0.08 & - \\
 V10 & EcB       & 0.242 & 13:27:42.00 & -47:07:24.15 & 20.29 & 1.46 & 0.90 & - \\
 V11 & EcB       & 0.286 & 13:27:39.29 & -47:09:26.64 & 20.36 & 1.57 & 0.70 & - \\
 V12 & AB RR Lyr & 0.683 & 13:27:32.83 & -47:13:43.38 & 14.62 & 0.60 & 0.90 & V149 \\
 V13 & EcB       & 0.287 & 13:27:29.62 & -47:13:09.65 & 17.42 & 1.03 & 0.50 & - \\
 V14 & EcB       & 0.920 & 13:27:59.82 & -47:13:12.57 & 16.40 & 1.07 & 0.05 & - \\
 V15 & Puls      & 0.405 & 13:27:22.35 & -47:11:12.86 & 15.92 & 0.94 & 0.01 & - \\
 V16 & LPV       & 23.79   & 13:27:15.73 & -47:13:10.13 & 16.16 & 1.07 & 0.09 & - \\
 V17 & EcB       & 0.398 & 13:27:02.82 & -47:08:49.08 & 16.34 & 0.43 & 0.35 & V213 \\
 V18 & AB RR Lyr     & 0.738 & 13:27:55.05 & -47:04:38.51 & 14.43 & 0.77 & 0.70 & V172 \\
 V19 & Puls/EcB     & 0.535 & 13:29:00.98 & -47:26:14.07 & 17.57 & 1.18 & 0.16 & - \\
 V20 & LPV       & 11.59 & 13:28:59.41 & -47:25:31.22 & 16.48 & 1.30 & 0.08 & - \\
 V21 & EcB       & 0.283 & 13:28:53.83 & -47:16:53.10 & 18.42 & 1.48 & 0.75 & - \\
 V22 & SX Phe  & 0.041 & 13:28:53.57 & -47:19:29.71 & 17.15 & 0.47 & 0.07 & - \\
 V23 & SX Phe  & 0.037 & 13:28:44.57 & -47:24:50.96 & 17.16 & 0.46 & 0.10 & - \\
 V24 & EcB       & 0.512 & 13:28:35.26 & -47:23:56.91 & 17.07 & 0.49 & 0.10 &  V255 \\
 V25 & EcB       & 0.404 & 13:28:32.81 & -47:26:24.32 & 17.71 & 0.95 & 0.55 & V245 \\
 V26 & EcB       & 0.426 & 13:28:25.94 & -47:23:52.31 & 14.56 & 0.73 & 0.09 & - \\
 V27 & LPV       & 20.3    & 13:28:25.03 & -47:16:41.47 & 15.08 & 1.35 & 0.09 & - \\
 V28 & EcB       & 0.311 & 13:28:18.70 & -47:18:34.48 & 19.80 & 0.85 & 0.50 & - \\
 V29 & LPV       & 18.09   & 13:28:23.48 & -47:21:16.62 & 13.95 & 1.08 & 0.05 & - \\
 V30 & EcB       & 0.258 & 13:29:09.27 & -47:24:31.15 & 20.96 & 1.64 & 0.55 & - \\
 V31 & LPV       & 23.25  & 13:27:43.09 & -47:26:35.59 & 15.82 & 1.13 & 0.05 & - \\
 V32 & C RR Lyr     & 0.336 & 13:27:35.61 & -47:26:30.58 & 14.45 & 0.53 & 0.40 & V82 \\
 V33 & C RR Lyr     & 0.299 & 13:27:30.20 & -47:28:05.37 & 14.65 & 0.49 & 0.40 & V19 \\
 V34 & EcB       & 0.332  & 13:27:28.71 & -47:26:19.71 & 16.38 & 0.49 & 0.12 & V240 \\
 V35 & EcB       & 0.385 & 13:27:28.68 & -47:27:39.28 & 16.38 & 0.41 & 0.11 & V254 \\
 V36 & AB RR Lyr     & 0.622 & 13:27:45.04 & -47:24:56.84 & 14.93 & 0.82 & 0.90 & V18 \\
 V37 & S.Det.EcB & 1.168 & 13:27:44.03 & -47:26:09.56 & 14.13 & 0.39 & 0.60 & V78 \\
 V38 & C RR Lyr     & 0.779 & 13:27:36.64 & -47:24:48.64 & 14.41 & 0.60 & 0.40 & V81 \\
 V39 & EcB       & 34.8$\ast$    & 13:27:42.06 & -47:23:34.95 & 15.45 & 1.04 & 0.12 & NV406 \\
 V40 & EcB       & 0.366 & 13:27:36.20 & -47:23:46.53 & 17.28 & 0.48 & 0.13 & V241 \\
 V41 & S.Det.EcB & 2.463 & 13:27:21.72 & -47:23:32.93 & 17.20 & 0.86 & 0.30 & V212 \\
 V42 & EcB       & 0.638 & 13:27:20.47 & -47:23:59.38 & 14.64 & 0.51 & 0.20 & V169 \\
 V43 & EcB       & 0.616 & 13:28:03.46 & -47:21:28.19 & 14.71 & 0.56 & 0.20 & V289 \\
 V44 & LPV       & 8.463   & 13:27:35.42 & -47:21:08.44 & 18.03 & 0.76 & 0.20 & - \\
 V45 & C RR Lyr     & 0.426 & 13:27:20.86 & -47:22:06.03 & 14.49 & 0.68 & 0.35 & V77 \\
 V46 & EcB       & 0.687   & 13:27:19.48 & -47:21:48.77 & 18.26 & 0.72 & 0.15 & - \\
 V47 & C RR Lyr     & 0.422 & 13:27:19.65 & -47:18:47.19 & 14.32 & 0.59 & 0.35 & V75 \\
 V48 & AB RR Lyr     & 0.503 & 13:27:07.22 & -47:17:34.43 & 15.11 & 0.91 & 1.10 & V74 \\
 V49 & C RR Lyr     & 0.315 & 13:29:04.07 & -47:36:21.78 & 14.76 & 0.57 & 0.30 & V177 \\
 V50 & Puls      & 0.162 & 13:29:00.54 & -47:30:23.80 & 19.66 & 1.16 & 0.20 & - \\
 V51 & LPV       & 3.102  & 13:29:00.27 & -47:36:37.57 & 18.39 & 2.07 & 0.10 & - \\
 V52 & LPV       & 4.337   & 13:28:59.84 & -47:36:12.18 & 16.14 & 1.10 & 0.05 & - \\
 V53 & C RR Lyr/EcB     & 0.377/0.754 & 13:28:54.73 & -47:30:21.08 & 19.41 & 0.89 & 0.25 & - \\
 V54 & S.Det.EcB & 1.901 & 13:28:53.72 & -47:37:35.04 & 15.86 & 0.52 & 0.50 & - \\
 V55 & LPV       & 9.259   & 13:28:51.95 & -47:38:07.15 & 18.66 & 0.75 & 0.40 & - \\
 V56 & EcB       & 0.284 & 13:28:37.20 & -47:34:28.95 & 18.95 & 1.00 & 0.24 & V244 \\
 V57 & EcB       & 0.743 & 13:29:05.37 & -47:39:49.94 & 16.88 & 0.97 & 0.07 & - \\
 V58 & LPV       & 9.608  & 13:29:07.01 & -47:37:32.15 & 15.93 & 0.86 & 0.02 & - \\
 V59 & Det.EcB   & 3.469 & 13:27:21.36 & -47:39:31.43 & 16.82 & 1.02 & 0.16 &  - \\
\noalign{\medskip}
\hline\hline
\end{tabular}
\end{center}
\vspace*{3mm}
\normalsize{The total variable star catalog. Tabulated are the identification
number of each variable, its type, period in days, J2000.0 equatorial coordinate
s, V magnitude and V-I color as intensity-averaged values measured from the CMD
dataset, the V+R amplitude, the alternate name from \citet{K2004} if the variabl
e is previously known. Those variables marked with a `$-$' in the Prev Name colu
mn are hence new discoveries. EcB refers to Eclipsing Binaries, with S.Det.EcB a
nd Det.EcB referring to semi-detached and detached systems respectively. LPV den
otes a long period variable. The periods marked with a $\ast$ denote those syste
ms which contain only one eclipse in the data and had their periods determined v
ia the visible secondary ellipsoidal variations.}

\newpage

\begin{center}
\footnotesize
\begin{tabular}{lllllllll} \hline\hline
\noalign{\medskip}
$ID$ & $Type$ & $Period (d)$ & $RA (J2000.0)$ & $DEC (J2000.0)$ & $V$ & $V-I$ &
$Prev Name$ \\
&&& h\hspace{2mm}m\hspace{2mm}s & $^{\circ}$\hspace{4mm}$'$\hspace{4mm}$''$ &&\\
\noalign{\medskip}
\hline
\noalign{\medskip}
 V60 & C RR Lyr     & 0.330 & 13:27:37.66 & -47:37:35.21 & 14.76 & 0.61 & 0.40 & V16 \\
 V61 & AB RR Lyr     & 0.794 & 13:27:49.36 & -47:36:50.78 & 14.51 & 0.70 & 0.50 & V57 \\
 V62 & EcB       & 0.342 & 13:27:21.77 & -47:37:19.43 & 16.64 & 0.34 & 0.16 & V214 \\
 V63 & C RR Lyr     & 0.396 & 13:27:41.01 & -47:34:08.19 & 14.61 & 0.70 & 0.40 & V22 \\
 V64 & C RR Lyr     & 0.462 & 13:27:38.29 & -47:34:15.02 & 14.39 & 0.70 & 0.30 & V24 \\
 V65 & SX Phe  & 0.047 & 13:27:53.90 & -47:31:54.39 & 16.92 & 0.30 & 0.40 & V194 \\
 V66 & C RR Lyr     & 0.335 & 13:27:45.99 & -47:32:44.37 & 14.73 & 0.60 & 0.40 & V105 \\
 V67 & C RR Lyr     & 0.391  & 13:27:27.73 & -47:33:43.18 & 14.33 & 0.55 & 0.30 & V70 \\
 V68 & EcB       & 0.241 & 13:27:22.95 & -47:32:18.83 & 17.28 & 0.89 & 0.04 & - \\
 V69 & LPV       & 10.64   & 13:27:08.59 & -47:32:54.86 & 16.23 & 0.99 & 0.02 & - \\
 V70 & AB RR Lyr     & 0.691 & 13:27:22.08 & -47:30:12.69 & 14.01 & -0.96 & 0.80 & V102 \\
 V71  & Det.EcB  & 6.110$\ast$  & 13:27:08.30 & -47:31:41.54 & 17.35 & 0.90 & 0.80 & - \\
 V72  & LPV       & 8.41    & 13:27:08.05 & -47:31:39.83 & 18.87 & 0.90 & 0.12 & - \\
 V73  & EcB       & 0.249 & 13:29:04.61 & -47:50:31.04 & 18.58 & 0.80 & 0.50 & - \\
 V74  & Puls   & 0.667 & 13:29:03.53 & -47:48:58.30 & 16.54 & 1.05 & 0.16 & - \\
 V75  & EcB       & 0.566 & 13:28:51.72 & -47:48:19.06 & 15.49 & 0.50 & 0.30 & - \\
 V76  & LPV       & 19.6    & 13:28:51.35 & -47:52:48.95 & 15.89 & 1.07 & 0.05 & - \\
 V77  & LPV       & 5.070  & 13:28:07.59 & -47:53:03.78 & 16.17 & 0.97 & 0.10 & - \\
 V78  & EcB       & 0.281 & 13:28:06.77 & -47:45:33.73 & 19.27 & 1.16 & 0.40 & V246 \\
 V79  & EcB       & 0.509 & 13:28:06.34 & -47:46:55.30 & 18.64 & 0.99 & 0.14 & - \\
 V80  & EcB       & 0.232 & 13:27:57.01 & -47:48:38.15 & 19.78 & 1.26 &  0.40& - \\
 V81  & LPV       & 2.733  & 13:27:38.59 & -47:42:51.23 & 16.93 & 1.01 & 0.05 & - \\
 V82  & AB RR Lyr  & 0.518 & 13:27:36.47 & -47:46:40.07 & 18.34 & 0.76 & 0.90 & V283 \\
 V83  & LPV       & 6.033   & 13:27:35.03 & -47:49:46.33 & 17.36 & 0.92  & 0.06 & - \\
 V84  & EcB       & 0.664 & 13:27:27.23 & -47:47:32.29 & 15.17 & 0.99 & 0.07 & V282 \\
 V85  & EcB       & 0.382 & 13:26:44.60 & -47:51:12.54 & 18.28 & 0.87 & 0.12 & - \\
 V86  & LPV       & 16.96   & 13:26:42.33 & -47:47:49.67 & 14.56 & 0.92 & 0.03 & - \\
 V87  & LPV       & 24.39   & 13:26:31.41 & -47:53:59.42 & 16.20 & 0.94 & 0.50 & - \\
 V88  & EcB       & 0.281 & 13:26:21.31 & -47:52:49.21 & 19.69 & 1.44 & 1.10 & - \\
 V89  & LPV       & 9.26    & 13:26:20.52 & -47:50:04.12 & 18.31 & 0.83 & 0.80 & - \\
 V90  & Det.EcB   & 1.634 & 13:26:18.70 & -47:53:14.44 & 18.72 & 1.39 & 0.90 &  - \\
 V91  & AB RR Lyr  & 0.638 & 13:26:14.03 & -47:53:08.18 & 17.12 & 0.77 & 0.35 & - \\
 V92  & SX Phe  & 0.099 & 13:26:07.49 & -47:47:16.85 & 19.73 & 1.11 & 0.15 & - \\
 V93  & LPV       & 6.289  & 13:25:04.23 & -47:50:52.73 & 13.97 & 0.86 & 0.02 & - \\
 V94  & EcB       & 0.337 & 13:25:03.82 & -47:49:32.58 & 17.81 & 0.83 & 0.50 & - \\
 V95  & EcB       & 0.472 & 13:24:59.90 & -47:48:32.27 & 19.99 & 2.14 & 0.20 &  - \\
 V96  & EcB       & 0.338 & 13:24:57.89 & -47:43:36.97 & 17.14 & 0.79 & 0.18 & - \\
 V97  & EcB       & 0.323 & 13:24:56.74 & -47:49:28.98 & 18.27 & 1.20 & 0.85 & - \\
 V98  & LPV       & 21.81  & 13:24:55.14 & -47:49:17.78 & 14.08 & 0.91 & 0.02 & - \\
 V99  & C RR Lyr/EcB  & 0.671/1.342 & 13:24:40.44 & -47:45:22.52 & 17.45 & 0.98 & 0.18 & - \\
 V100 & Det.EcB   & 9.232  & 13:24:36.22 & -47:48:37.29 & 18.87 & 1.20 &  0.70 & - \\
 V101 & AB RR Lyr        & 0.605 & 13:26:07.69 & -47:37:55.74 & 14.66 & 0.77  & 0.80 & V49 \\
 V102 & Det.EcB   & 1.496 & 13:26:47.32 & -47:35:58.83 & 16.91 & 0.51 & 0.50 & V210 \\
 V103 & C RR Lyr        & 0.344 & 13:26:02.12 & -47:36:19.50 & 14.39 & 0.45 & 0.40 & V64 \\
 V104 & LPV       & 9.626  & 13:25:49.68 & -47:37:00.26 & 17.05 & 1.33 & 0.30 & NV384 \\
 V105 & AB RR Lyr        & 0.631 & 13:26:12.24 & -47:34:17.44 & 14.00 & 0.42 & 0.90 & V115 \\
 V106 & AB RR Lyr        & 0.567 & 13:26:22.35 & -47:34:34.98 & 14.27 & 0.52 & 0.80 & V44 \\
 V107 & AB RR Lyr        & 0.734 & 13:26:07.14 & -47:33:10.35 & 14.30 & 0.63 & 0.70 &  V34 \\
 V108 & SX Phe   & 0.041 & 13:26:40.49 & -47:33:45.12 & 17.27 & 0.54 & 0.09 & V228 \\
 V109 & TypeII Ceph   & 1.349 & 13:26:35.69 & -47:32:47.03 & 13.22 & 0.98 & 0.82 & V60 \\
 V110 & AB RR Lyr        & 0.635 & 13:26:30.29 & -47:33:01.40 & 14.50 & 0.71 & 0.90 &  V122 \\
 V111 & AB RR Lyr        & 0.548 & 13:26:25.48 & -47:32:47.87 & 14.76 & 0.82 & 0.80 & V120 \\
 V112 & C RR Lyr        & 0.367 & 13:26:45.33 & -47:30:37.99 & 14.21 & 0.75 & 0.22 & V158 \\
 V113 & EcB       & 0.295 & 13:26:41.76 & -47:31:28.91 & 16.81 & 1.00  & 0.45 & NV338 \\
 V114 & AB RR Lyr        & 0.611 & 13:26:40.55 & -47:30:17.31 & 14.35 & 0.65 & 0.70 & V118 \\
 V115 & C RR Lyr       & 0.321 & 13:26:39.64 & -47:30:26.74 & 14.92 & 0.95 & 0.24 & V48 \\
 V116 & C RR Lyr       & 0.306 & 13:26:38.28 & -47:31:16.53 & 14.74 & 0.59 & 0.20 & V119 \\
 V117 & C RR Lyr       & 0.304 & 13:26:28.15 & -47:31:49.74 & 14.83 & 0.42 & 0.40 & V121 \\
 V118 & AB RR Lyr        & 0.634 & 13:26:24.52 & -47:30:45.27 & 13.92 & 0.43 & 0.91 & V40 \\
 V119 & SX Phe   & 0.047 & 13:26:20.44 & -47:31:58.92 & 17.09 & 0.60 & 0.19 & V197 \\
\noalign{\medskip}
\hline\hline
\end{tabular}
\end{center}
\vspace*{3mm}
\normalsize{Table 2 (continued).}

\newpage

\begin{center}
\footnotesize
\begin{tabular}{lllllllll} \hline\hline
\noalign{\medskip}
$ID$ & $Type$ & $Period (d)$ & $RA (J2000.0)$ & $DEC (J2000.0)$ & $V$ & $V-I$ &
$Prev Name$ \\
&&& h\hspace{2mm}m\hspace{2mm}s & $^{\circ}$\hspace{4mm}$'$\hspace{4mm}$''$ &&\\
\noalign{\medskip}
\hline
\noalign{\medskip}
 V120 & SX Phe   & 0.051 & 13:26:18.07 & -47:30:34.81 & 16.05 & 0.90 & 0.24 & NV324 \\
 V121 & EcB       & 0.596 & 13:26:17.72 & -47:30:22.77 & 14.57 & 0.53 & 0.06 & NV357 \\
 V122 & AB RR Lyr        & 0.835  & 13:26:17.57 & -47:30:10.93 & 14.59 & 1.60 & 0.65 & V128 \\
 V123 & LPV       & 23.35   & 13:26:08.18 & -47:30:31.79 & 15.12 & 1.09 & 0.07 & V216 \\
 V124 & EcB       & 0.248 & 13:26:08.18 & -47:31:26.32 & 17.79 & 0.73 & 0.14 & NV332 \\
 V125 & LPV       & 4.808  & 13:25:31.89 & -47:31:55.63 & 14.61  & 0.81 & 0.04 & - \\
 V126 & LPV       & 23.32   & 13:25:21.47 & -47:34:06.77 & 16.90 & 1.10 & 0.03 & - \\
 V127 & AB RR Lyr        & 0.826 & 13:25:07.87 & -47:36:54.11 & 14.61 & 0.81 & 0.30 & V63 \\
 V128 & AB RR Lyr        & 0.653 & 13:25:10.93 & -47:37:33.52 & 14.65 & 0.74 & 0.90 & V69 \\
 V129 & LPV       & 3.336  & 13:24:58.54 & -47:36:07.77 & 18.53 & 1.41 & 0.18 & V284 \\
 V130 & LPV       & 9.815   & 13:24:56.93 & -47:32:28.83 & 14.04 & 1.04 & 0.04 & - \\
 V131 & SX Phe   & 0.047 & 13:24:54.09 & -47:41:03.53 & 17.24 & 0.73 & 0.14 &  - \\
 V132 & EcB       & 0.341 & 13:24:49.58 & -47:33:30.42 & 17.22 & 1.02 & 0.04 & - \\
 V133 & AB RR Lyr        & 0.844 & 13:26:41.09 & -47:28:18.74 & 14.47 & 0.58 & 0.30 & V144 \\
 V134 & C RR Lyr        & 0.316 & 13:26:40.45 & -47:26:36.85 & 15.25 & 1.43 & 0.20 & V267 \\
 V135 & C RR Lyr        & 0.352 & 13:26:39.93 & -47:28:03.81 & 14.52 &	0.58 & 0.24 & NV356 \\
 V136 & AB RR Lyr        & 0.500 & 13:26:39.66 & -47:26:55.80 & 14.50 & 0.26 & 0.75 & V165 \\
 V137 & AB RR Lyr        & 0.624 & 13:26:39.51 & -47:27:04.97 & 14.16  & 0.44 & 0.07 & V96 \\
 V138 & AB RR Lyr        & 0.677 & 13:26:37.99 & -47:27:36.86 & 14.10 &	0.68 & 0.80 & V139 \\
 V139 & AB RR Lyr        & 0.661 & 13:26:35.42 & -47:28:05.63 & 13.97 &	0.61 & 0.55 & V52 \\
 V140 & C RR Lyr        & 0.334 & 13:26:31.72 & -47:27:05.67 & 14.22 & 0.26 & 0.45 & V137 \\
 V141 & LPV       & 14.48  & 13:26:31.62 & -47:27:26.01 & 14.04 & 0.99 & 0.07 & NV386 \\
 V142 & AB RR Lyr     & 0.619 & 13:26:26.77 & -47:27:57.00 & 14.17 & 0.54 & 1.00 & V62 \\
 V143 & AB RR Lyr     & 0.615 & 13:26:26.22 & -47:28:17.93 & 15.48 & 1.22 & 0.45 & V27 \\
 V144 & AB RR Lyr     & 0.312  & 13:26:43.18 & -47:25:58.06 & 14.66 & 0.50 & 0.20 & - \\
 V145 & AB RR Lyr     & 0.811 & 13:26:27.16 & -47:24:47.36 & 16.01 & 2.11 & 0.75 & - \\
 V146 & AB RR Lyr$?$ & 1.009$?$ & 13:26:13.21 & -47:26:10.97 & 14.28 & 0.75 & $\ge$0.20 & V263 \\
 V147 & C RR Lyr     & 0.381 & 13:26:11.24 & -47:26:00.03 & 14.41 & -0.31 & 0.40 & V21 \\
 V148 & C RR Lyr     & 0.311 & 13:26:43.92 & -47:22:49.09 & 14.58 & 0.51 & 0.24 & V274 \\
 V149 & AB RR Lyr     & 0.574 & 13:26:42.81 & -47:24:22.63 & 13.83 & 0.34 & 0.90 & V51 \\
 V150 & S.Det.EcB & 0.369 & 13:26:33.44 & -47:23:01.03 & 17.23 & 0.53 & 0.80 & V205 \\
 V151 & C RR Lyr     & 0.387 & 13:26:27.29 & -47:24:07.39 & 14.69 & 0.66 & 0.40 & V12 \\
 V152 & AB RR Lyr     & 0.515 & 13:26:18.39 & -47:23:13.33 & 14.76 & 0.67 & 1.00 & V5 \\
 V153 & Det.EcB   & 0.834 & 13:26:17.40 & -47:23:17.76 & 16.58 & 0.19 & 0.32 & V209 \\
 V154 & SX Phe & 0.038 & 13:26:14.44 & -47:23:54.75 & 17.36 & 0.49 & 0.08 & V227 \\
 V155 & C RR Lyr/EcB     & 0.370/0.740 & 13:26:13.17 & -47:24:04.54 & 16.10 & -0.15 & 0.20 & V58 \\
 V156 & AB RR Lyr     & 0.628 & 13:26:12.99 & -47:24:19.81 & 14.76 & 0.77 & 1.00 & V4 \\
 V157 & LPV       & 22.3    & 13:26:12.20 & -47:24:37.74 & 17.52 & 1.31 & $\ge$0.65 & - \\
 V158 & C RR Lyr     & 0.375 & 13:26:07.04 & -47:24:37.49 & 14.29 & 0.48 & 0.35 & V10 \\
 V159 & LPV       & 23.82   & 13:26:05.67 & -47:23:54.47 & 17.93 & 0.84 & 0.80 & - \\
 V160 & LPV       & 25.11   & 13:26:03.80 & -47:23:36.83 & 18.61 & 0.81 & 0.90 & - \\
 V161 & Det.EcB  & 3.697$\ast$  & 13:26:38.87 & -47:21:40.61 & 16.22 & 0.99 & 0.22 & NV409 \\
 V162 & C RR Lyr     & 0.407 & 13:26:33.17 & -47:22:26.01 & 13.82 & -0.03 & 0.40 & V66 \\
 V163 & S.Det.EcB & 0.823 & 13:26:22.79 & -47:21:43.44 & 17.85 & 0.84 & 0.35 & NV363 \\
 V164 & TypeII Ceph     & 1.346 & 13:26:14.86 & -47:21:15.17 & 14.09 & 0.82 & 0.27 & V92 \\
 V165 & C RR Lyr     & 0.333 & 13:26:04.07 & -47:21:47.22 & 14.54 & 0.53 & 0.16 & V185 \\
 V166 & EcB       & 2.059 & 13:26:38.55 & -47:20:00.09 & 14.67 & 0.82 & 0.20 & NV378 \\
 V167 & AB RR Lyr     & 0.535 & 13:26:12.79 & -47:19:36.27 & 14.45 & 0.68 & 0.35 & V68 \\
 V168 & SX Phe & 0.039 & 13:26:08.40 & -47:19:24.65 & 17.15 & 0.52 & 0.09 & V219 \\
 V169 & C RR Lyr     & 0.313 & 13:25:49.44 & -47:20:21.88 & 14.56 & 0.46 & 0.20 & V163 \\
 V170 & AB RR Lyr     & 0.564  & 13:26:28.60 & -47:18:47.39 & 14.95 & 0.73 & 0.80 & V67 \\
 V171 & AB RR Lyr     & 0.773 & 13:26:23.53 & -47:18:48.25 & 14.45 & -1.42 & 0.50 & V54 \\
 V172 & SX Phe & 0.044 & 13:26:11.19 & -47:17:54.05 & 17.08 & 0.50 & 0.08 & V218 \\
 V173 & AB RR Lyr     & 0.589 & 13:25:30.83 & -47:27:21.21 & 14.92 & 0.78 & 1.00 & V45 \\
 V174 & AB RR Lyr     & 0.687 & 13:25:30.21 & -47:25:51.91 & 14.71 & 0.79 & 0.80 & V46 \\
 V175 & SX Phe & 0.057 & 13:25:01.16 & -47:25:29.69 & 17.14 & 0.58 & 0.20 & V196 \\
 V176 & AB RR Lyr     & 0.743 & 13:25:06.56 & -47:23:33.70 & 14.47 & 0.73 & 0.60 & V85 \\
 V177 & EcB       & 0.416 & 13:25:02.08 & -47:24:15.45 & 16.86 & 0.59 & 0.09 & V248 \\
 V178 & EcB       & 1.094  & 13:25:19.22 & -47:22:31.05 & 18.44 & 1.01 & 0.30 & V235 \\
 V179 & LPV       & 17.85  & 13:25:00.42 & -47:17:39.99 & 18.38 & 1.24 & 0.35 & - \\
 V180 & SX Phe & 0.046 & 13:26:38.93 & -47:11:51.28 & 17.10 & 0.59 & 0.13 & V202 \\
 V181 & S.Det.EcB & 0.669 & 13:26:20.53 & -47:04:27.64 & 16.84 & 1.00 & 1.10 & - \\
 V182 & SX Phe & 0.037 & 13:26:17.27 & -47:11:49.95 & 17.52 & 0.57 & 0.05 & V232 \\
 V183 & AB RR Lyr     & 0.493 & 13:26:09.94 & -47:13:40.04 & 15.19 & 0.84 & 0.80 & V130 \\
 V184 & EcB       & 0.406 & 13:25:15.77 & -47:03:54.87 & 17.33 & 0.90 & 0.07 & - \\
 V185 & AB RR Lyr     & 0.653 & 13:25:13.28 & -47:12:28.64 & 14.77 & 0.78 & 0.80 & V134 \\
 V186 & EcB       & 0.311 & 13:24:45.29 & -47:09:17.90 & 19.40 & 0.96 & 0.20 & - \\
 V187 & LPV       & 8.63   & 13:24:36.96 & -47:07:32.53 & 14.88 & 0.95 & 0.03 & - \\
\noalign{\medskip}
\hline\hline
\end{tabular}
\end{center}
\vspace*{3mm}
\normalsize{Table 2 (continued).}


\end{document}